\documentclass[conference]{IEEEtran}
\usepackage{color,xcolor}
\usepackage{color,soul}
\usepackage{subfigure}
\usepackage{cleveref}
\usepackage{balance}
\usepackage{graphicx}
\usepackage{url}
\usepackage{multirow}
\usepackage{graphicx}
\usepackage{amssymb}
\usepackage{amsfonts}
\usepackage[small,bf]{caption}
\usepackage{wrapfig}
\usepackage{alltt}
\usepackage{booktabs}
\usepackage{xspace}
\usepackage{verbatim}
\usepackage{footnote}
\usepackage{epstopdf}
\definecolor{darkgreen}{RGB}{47,109,79}
\definecolor{darkblue}{RGB}{57,79,99}

\usepackage[keeplastbox]{flushend}

\newcommand{\precaption}{\vspace{0cm}}
\newcommand{\postcaption}{\vspace{0cm}}

\newcommand{\descr}[1]{\medskip\noindent\textbf{#1}}

\usepackage[marginal,hang,stable]{footmisc}
\renewcommand{\footnoterule}{%
  \kern -3pt
  \hrule width 1in
  \kern 2pt
}

\makeatletter
\def\url@foostyle{%
  \@ifundefined{selectfont}{\def\UrlFont{\rm}}{\def\UrlFont{\rmfamily}}}
\makeatother
\renewcommand{\footnotesize}{\fontsize{8}{9}\selectfont}

\ifCLASSOPTIONcompsoc
  \usepackage[nocompress]{cite}
\else
  \usepackage{cite}
\fi

\makeatletter
\def\@copyrightspace{\relax}
\makeatother
\begin{document} 
\pagenumbering{arabic}

\title{Characterizing Key Stakeholders in an \\Online Black-Hat Marketplace}

\author{\IEEEauthorblockN{Shehroze Farooqi}
\IEEEauthorblockA{University of Iowa}
\IEEEauthorblockA{Iowa City, IA, USA}
\IEEEauthorblockA{shehroze-farooqi@uiowa.edu}\\
\IEEEauthorblockN{Guillaume Jourjon}
\IEEEauthorblockA{Data61 CSIRO}
\IEEEauthorblockA{Sydney, NSW, Australia}
\IEEEauthorblockA{Guillaume.Jourjon@data61.csiro.au}
\and
\IEEEauthorblockN{Muhammad Ikram}
\IEEEauthorblockA{Data61 CSIRO, UNSW}
\IEEEauthorblockA{Sydney, NSW, Australia}
\IEEEauthorblockA{Muhammad.Ikram@data61.csiro.au}\\
\IEEEauthorblockN{Mohamed Ali Kaafar}
\IEEEauthorblockA{Data61 CSIRO}
\IEEEauthorblockA{Sydney, NSW, Australia}
\IEEEauthorblockA{dali.kaafar@data61.csiro.au}
\and
\IEEEauthorblockN{Emiliano De Cristofaro}
\IEEEauthorblockA{University College London}
\IEEEauthorblockA{London, UK}
\IEEEauthorblockA{e.decristofaro@ucl.ac.uk}\\
\IEEEauthorblockN{Zubair Shafiq}
\IEEEauthorblockA{University of Iowa}
\IEEEauthorblockA{Iowa City, IA, USA}
\IEEEauthorblockA{zubair-shafiq@uiowa.edu}
\and
\IEEEauthorblockN{Arik Friedman}
\IEEEauthorblockA{Atlassian}
\IEEEauthorblockA{Sydney, NSW, Australia}
\IEEEauthorblockA{arik.friedman@gmail.com}\\
\IEEEauthorblockN{Fareed Zaffar}
\IEEEauthorblockA{LUMS}
\IEEEauthorblockA{Lahore, Pakistan}
\IEEEauthorblockA{fareed.zaffar@lums.edu.pk}

}

\maketitle

\begin{abstract}
Over the past few years, many black-hat marketplaces have emerged that facilitate access to reputation manipulation services such as fake Facebook likes, fraudulent search engine optimization (SEO), or bogus Amazon reviews.
In order to deploy effective technical and legal countermeasures, it is important to understand how these black-hat marketplaces operate, shedding light on the services they offer, who is selling, who is buying, what are they buying, who is more successful, why are they successful, etc.
Toward this goal, in this paper, we present a detailed micro-economic analysis of a popular online black-hat marketplace, namely, \url{SEOClerks.com}.
As the site provides non-anonymized transaction information, we set to analyze selling and buying behavior of individual users, propose a strategy to identify \emph{key users}, and study their tactics as compared to other (\emph{non-key}) users.
We find that \emph{key users}: (1) are mostly located in Asian countries, (2) are focused more on selling black-hat SEO services, (3) tend to list more lower priced services, and (4) sometimes buy services from other sellers and then sell at higher prices.
Finally, we discuss the implications of our analysis with respect to devising effective economic and legal intervention strategies against marketplace operators and \emph{key users}.
\end{abstract}

\section{Introduction} \label{sec: introduction}

Reputation plays a very important role in online services including e-commerce sites, search engines, or online social networks.
For instance, Amazon uses customer reviews to help users assess the credibility of sellers, Google relies on PageRank to determine search ranking of websites, while Facebook likes often offer a measure of the popularity of brands.
As a result, it is not surprising that an increasing number of black-hat marketplaces facilitate access to reputation manipulation services.
A multitude of \textit{online} and \textit{underground} (i.e., hosted as Tor hidden services) black-hat marketplaces sell services to generate bogus reviews, obtain fake likes, artificially boost PageRank, etc.
Several companies such as Amazon and Facebook have filed lawsuits against users who provide reputation manipulation services \cite{facebooksuit,amazonsuit}.
For instance, Amazon recently conducted a sting operation on Fiverr and sued more than a thousand ``John Doe'' fraudsters for selling bogus reviews  \cite{amazonlawsuit}.
Law enforcement agencies have also cracked down on different underground black-hat marketplaces \cite{silkRoadCrackDown,silkRoad2CrackDown,silkRoadDutchCrackDown}.

However, the cleanup or closure of a black-hat marketplace typically leads to increased popularity of other services~\cite{Soska2015MeasuringAnonymousUSENIX}.
In a way, the overall black-hat marketplace ecosystem is generally robust to such measures, highlighting the multifaceted and complex nature of the problem.
Therefore, the design and implementation of effective technical and legal countermeasures requires a thorough examination and deep understanding of how these black-hat marketplaces operate.
Prior work has studied their evolution and the types of fraudulent and illicit services they offer~ \cite{wang12crowdturfingWWW,motoyama11dirtyjobs,motoyama11undergroundforums,christin13silkroad,lee14darksidemicrotask,lee15turfingfivertwitter,levchenko11clicktrajectories,mccoy12priceless,mccoy12pharmaleaks,stringhini12twitterfollowermarketWOSN,thomas13traffickingfraudtwitteraccounts,stringhini13twitterfollower,Soska2015MeasuringAnonymousUSENIX,Xu:2015:ERM:}.
However, very little work has focused on individual sellers, buyers, and services: arguably, such an analysis is quite challenging, as most online and underground marketplaces do not reveal detailed buyer-seller transaction information. For instance, many black-hat marketplaces only provide aggregate positive and negative ratings which makes it impossible to track specific transactions among users on the marketplace.

Aiming to address this gap, this paper presents a first-of-its-kind, detailed micro-economic analysis of a popular online black-hat marketplace: \url{SEOClerks.com}.
We select SEOClerks as it provides detailed ratings, allowing us to analyze individual transaction-level information.
Moreover, SEOClerks is more popular than most of the other online black-hat marketplaces studied in prior work (e.g.,~\cite{wang12crowdturfingWWW,Xu:2015:ERM:}).
At the time of writing, SEOClerks is ranked in the top 12K websites globally by Alexa; whereas, for example, \url{Sandaha.com} is ranked 213K, \url{Zhubajei.com} 353K, and \url{Shuakewang.com} 1,128K.

Our goal is to identify key stakeholders on online black-hat marketplaces and understand their role in order to develop effective countermeasures.
First, we identify \emph{key users} who are among the early joiners, are very active, and make the most money on the marketplace.
Next, we characterize how key users differ as compared to other (\emph{non-key}) users.
We compare and contrast key and non-key users in terms of the services they offer, and their selling and buying behavior.

We start our analysis with a general characterization of SEOClerks, finding that it has over 262K users and 39K listed services. Using individual buyer ratings as a proxy for sales, our lower-bound estimate of the marketplace revenue is \$1.3 million. Moreover, we estimate that SEOClerks operators have earned hundreds of thousands of dollars from fees/commissions and advertising.

Next, we look for key users on the marketplace, identifying 99 of them. These are among the early joiners (accounts were registered around the launch of the marketplace), are very active (they have logged on to the site within a week of our crawl), and/or make the most money on the marketplace. These users, although accounting for less than 0.04\% of all users
and offering only 9\% of all services, actually generate 56\% of the marketplace revenue. We also find that a majority of key users are located in Asian countries (India, Indonesia), while buyers are relatively concentrated in European and North American countries (USA, UK, Italy).
The vast majority of services on SEOClerks are fraudulent, e.g., selling inbound links from other web pages (``backlinks'') to improve Google PageRank, inflating website traffic for click fraud, fake Instagram followers, Twitter retweets, or Facebook likes. Black-hat SEO services offered by key users actually account for a majority of their revenue. Key users are typically allowed to offer lower priced services (starting at \$1) and their services tend to receive more views than the services offered by other users.

Also, some key users purchase services from other sellers on SEOClerks and sell it at higher prices. For example, a key user offers a service for bogus SoundCloud plays and has also repeatedly purchased a similar service from another seller.
Finally, we show that SEOClerks operators use an escrow mechanism to get transaction/commission fees and to resolve disputes between sellers and buyers; thus, their marketplace accounts on PayPal, Payza, and BitPay can be targeted for economic and legal intervention.

Overall, black-hat marketplaces constitute a key link in the Internet fraud chain \cite{levchenko11clicktrajectories}.
Through their characterization, our work aims to help in devising effective economic and legal intervention strategies.
Since key users constitute a majority of the marketplace revenue, targeting them can considerably limit fraudulent activities out of black-hat marketplaces.
\section{Data} \label{sec: dataset}

\descr{Data Collection.} We conducted a complete crawl of \url{SEOClerks.com} in February 2015, using the Scrapy web crawler\footnote{\url{http://www.scrapy.org}}.
SEOClerks has a directory of user profiles that contains username, account creation date, last login date, location, user reputation level, average response time, ratings, description of skills, and the list of services offered.
SEOClerks also has a directory of services that contains service price, service creation date, a description of the service, seller's username, expected delivery time, number of orders in progress, number of views, and positive/negative buyer ratings.
We collected all publicly available information from both {\em user} and {\em service} directories.
We also crawled individual buyer ratings on service pages to identify their buyers.

\begin{table}[!t]
\centering
\begin{tabular}{|l|r|}
\hline
{Number of Users} & 262,909\\
\hline
{Number of Services} & 39,520\\
\hline
{Number of Services Sold} & 8,862\\
\hline
{Total Revenue} & \$1,349,316\\
\hline
{Average Revenue per service} &  \$152\\
\hline
{Alexa Global Rank} & 12K\\
\hline
\end{tabular}
\caption{Statistics of SEOClerks marketplace.}
\label{tbl:marketplaces}
\end{table}

\descr{General Statistics.}
Table~\ref{tbl:marketplaces} summarizes overall statistics of the SEOClerks marketplace.
SEOClerks is ranked by Alexa in the top 12K websites globally and top 3K in India.
Our crawled data includes 262,909 users and 39,520 services.
22\% of the services on SEOClerks are sold at least once.
The average revenue per sold service is \$152.
The estimated total revenue of SEOClerks is \$1,349,316, which is obtained by multiplying the price of each service with the corresponding rating count.
Since buyers are not required but are highly-recommended to rate the purchased services, our estimate represents a lower-bound on the actual total revenue.
We also note that several services include some add-ons (or ``service extras'') for additional payment.
From our crawls, we cannot identify the purchase of these add-ons.
Thus, our lower bound on the estimated revenue does not include service extras.

\descr{Ethical Considerations.}
As we collected and analyzed data pertaining to possibly fraudulent activities, we requested approval from our Institutional Review Board, which classified our research as {\em exempt}.
We note that: (1) we did not engage in any fraudulent transactions at the marketplace, and (2) we only collected publicly available information. %

\section{Identifying Key Stakeholders} \label{sec: Insider accounts}
Our work aims to identify and analyze key stakeholders who are crucial for the success of a black-hat marketplace.
We hypothesize that \emph{key users} of an online black-hat marketplace (1) join the marketplace soon after it was launched; (2) are among the most successful sellers on the marketplace; and (3) are very active on the marketplace.
Below, we further discuss and use these three criteria to identify key users on SEOClerks.

\begin{figure*}[t]
\centering
    \subfigure[Registration of Users over time]{
    \includegraphics[width=.95\columnwidth]{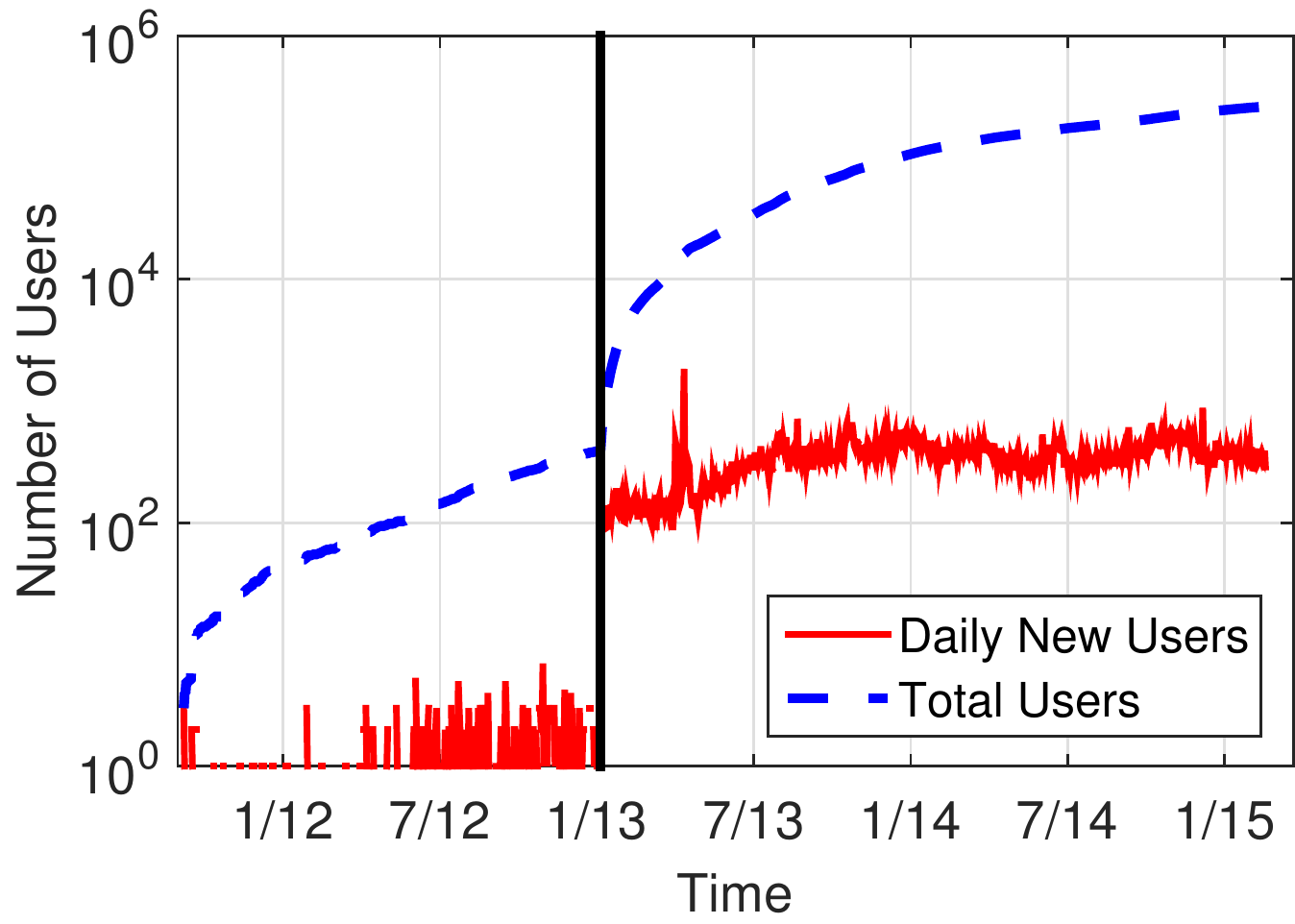}
    \label{fig: evolution}
    }
    \subfigure[Distribution of Sellers' Revenue]{
    \includegraphics[width=.95\columnwidth]{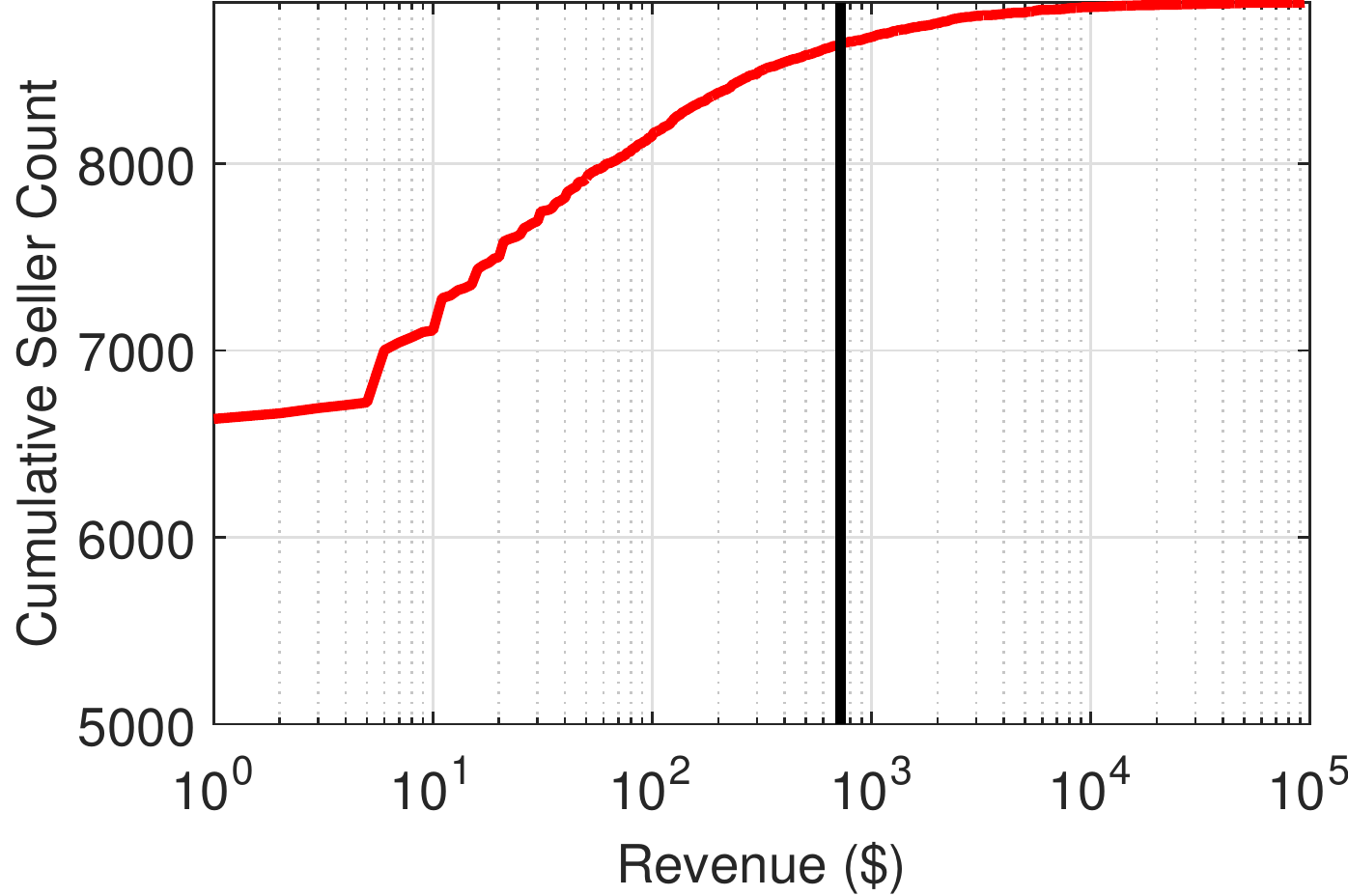}
    \label{fig: revenue cutoff}
    }
    \subfigure[Distribution of Last Login Date of Sellers.]{
    \includegraphics[width=.95\columnwidth]{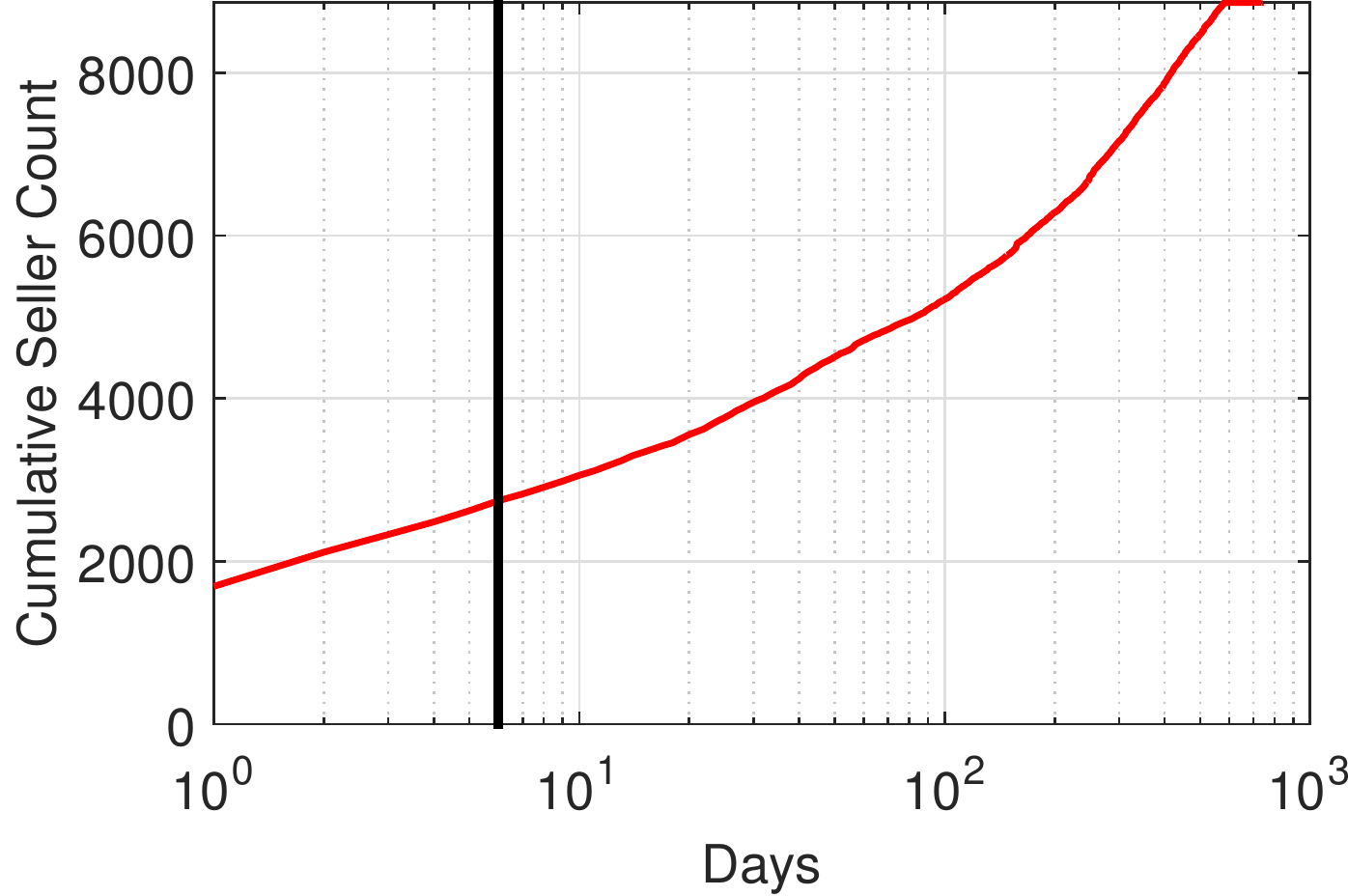}
    \label{fig: last login}
    }
    \subfigure[Relationship between seller join date, last login date, and revenue. Circle size represents seller revenue. Red circles represent key users while blue circles represent non-key users.]{
    \includegraphics[width=.95\columnwidth]{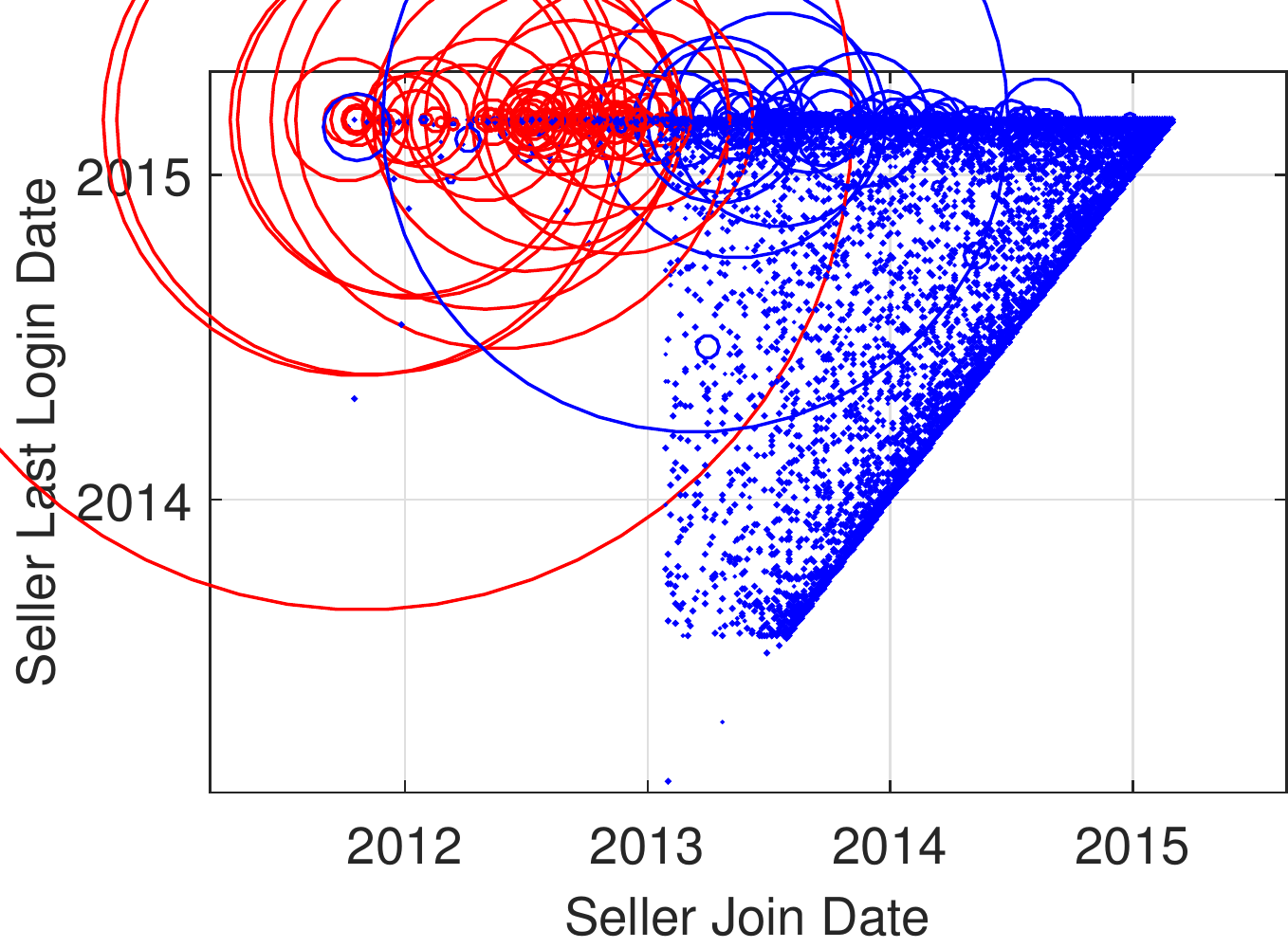}
    \label{fig: scatter join login}
    }
\precaption
\caption{Identification of \emph{key users} on SEOClerks}
\label{fig: insider Ring}
\postcaption
\end{figure*}

\descr{Early Joiners.}
We first analyze the registration of users over time  on SEOClerks using the account creation date reported for each user.
Figure \ref{fig: evolution} plots the daily registration rate of new users and the cumulative number of users on SEOClerks.
We note that the first user account was registered in mid-2011.
Our assessment is confirmed by the Internet Archive Wayback Machine\footnote{\url{http://web.archive.org}}, which has the first snapshot of SEOClerks dating back to October 7, 2011.
Note that the number of users initially grew fairly slowly (daily new users $<$ 10).
The marketplace experienced a sudden increase in new users beginning early 2013.
The increase in the number of new users might be explained by an aggressive social media campaign in early 2013 (offering \$2 promotional credit for tweeting about SEOClerks)\footnote{\url{http://web.archive.org/web/20130102230631/http://www.seoclerks.com/freemoney}}.
The vertical black line in Figure \ref{fig: evolution} marks the change point in early 2013 after which we observe a sharp increase in new user registrations.
The users who joined the marketplace before this cutoff date are labeled as \emph{early joiners}. Using this criterion, we identify a total of 391 early joiners.

\descr{Top Sellers.}
We define a user as a seller if the user has posted at least one service on SEOClerks.
In total, we identify 8,861 sellers on SEOClerks.
Figure \ref{fig: revenue cutoff} plots the revenue distribution for sellers on SEOClerks.
Out of 8,861 sellers, only 2,228 sellers sold at least one service.
The long-tail distribution indicates that a small number of sellers account for most of the marketplace revenue.
We label the top 10\% sellers (marked by the vertical black line) among the 2,228 sellers as \emph{top sellers}.
These 222 top sellers account for \$1,181,339 (88\%) revenue on SEOClerks.

\descr{Active Sellers.}
We identify active sellers on SEOClerks by analyzing their last login date.
Whenever a user logs in to SEOClerks, the last login date is updated on the user's profile.
Figure \ref{fig: last login} plots the distribution of sellers' last login date (at the time of our crawl) on SEOClerks.
We observe that more than half of the sellers on SEOClerks are not active.
We note that 2,826 (32\%) sellers logged in to the marketplace within a week of our crawl (marked by the vertical black line).
Since active sellers need to log in frequently in order to respond to customers and receive new orders, we label these 2,826 sellers who logged in within a week of our crawl as \emph{active sellers}.

\descr{Identifying Key Users.}
Figure \ref{fig: scatter join login} visualizes marketplace sellers using a scatter plot for join date and last login date, where the radius of each circle is proportional to the seller revenue.
We mark the users who satisfy the aforementioned three criteria with red circles.
The remaining users are marked with blue circles.
It is surprising to note that a vast majority of users who joined the marketplace before 2013 logged in very recently.
We also observe that a majority of these users are also top sellers on SEOClerks.
We label a total of 99 sellers who satisfy the aforementioned three criteria as \emph{\textbf{key users}}.
We next analyze the characteristics of these key users with the aim of facilitating the design of technical countermeasures and strategies for economic or legal intervention.

\section{Marketplace Analysis} \label{sec: characterization}
This section presents an in-depth analysis of SEOClerks with an emphasis on comparing and contrasting key users and non-key users.
We analyze a wide range of characteristics for services, sellers, and buyers on the marketplace.

\subsection{Services}
A vast majority of services on SEOClerks are geared towards fraudulent services such as selling backlinks for black-hat SEO, website traffic, Instagram followers, Twitter retweets, Facebook likes, URL spam, etc.
We identified a total of 39,520 services offered on SEOClerks.
A total of 3,645 (9\%) services were posted by key users, while the remaining 35,875 (91\%) services were posted by non-key users.
Below we characterize different aspects of the services offered by key users and non-key users.

\begin{figure*}[!t]
\centering
    \subfigure[Price]{
    \includegraphics[width=.95\columnwidth]{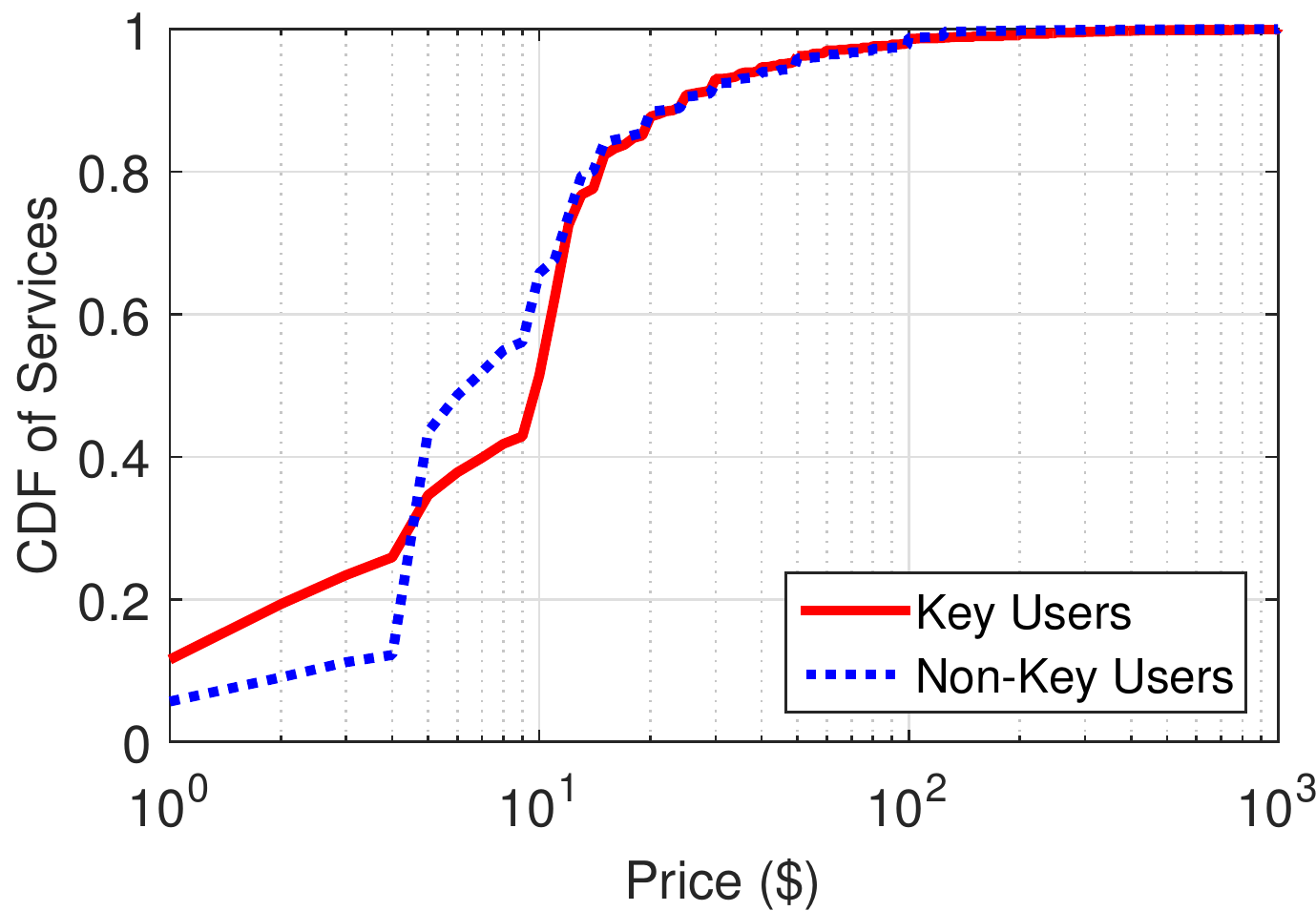}
    \label{fig: service price}
    }
    \subfigure[Volume]{
    \includegraphics[width=.95\columnwidth]{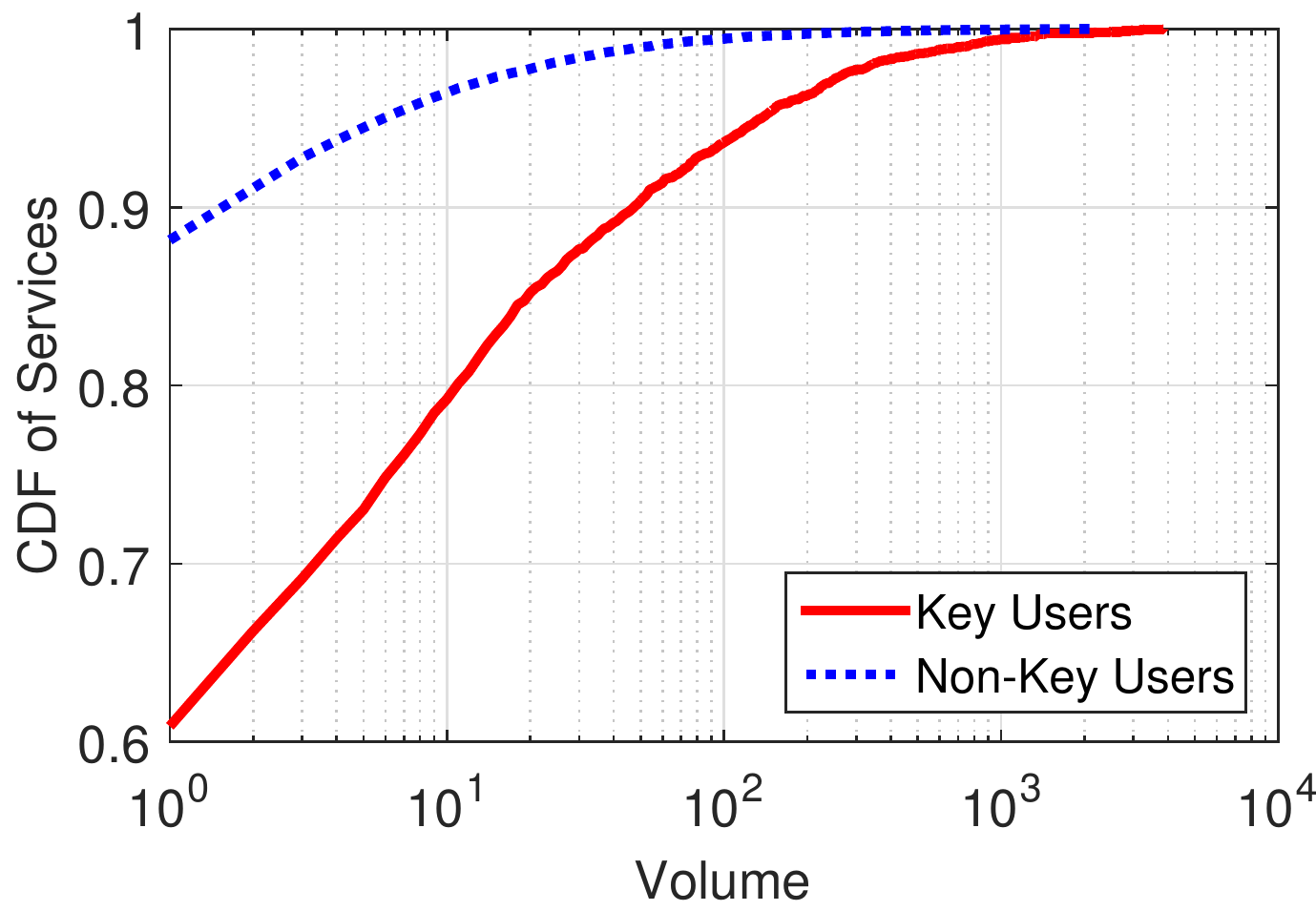}
    \label{fig: service volume}
    }
    \subfigure[Revenue]{
    \includegraphics[width=.95\columnwidth]{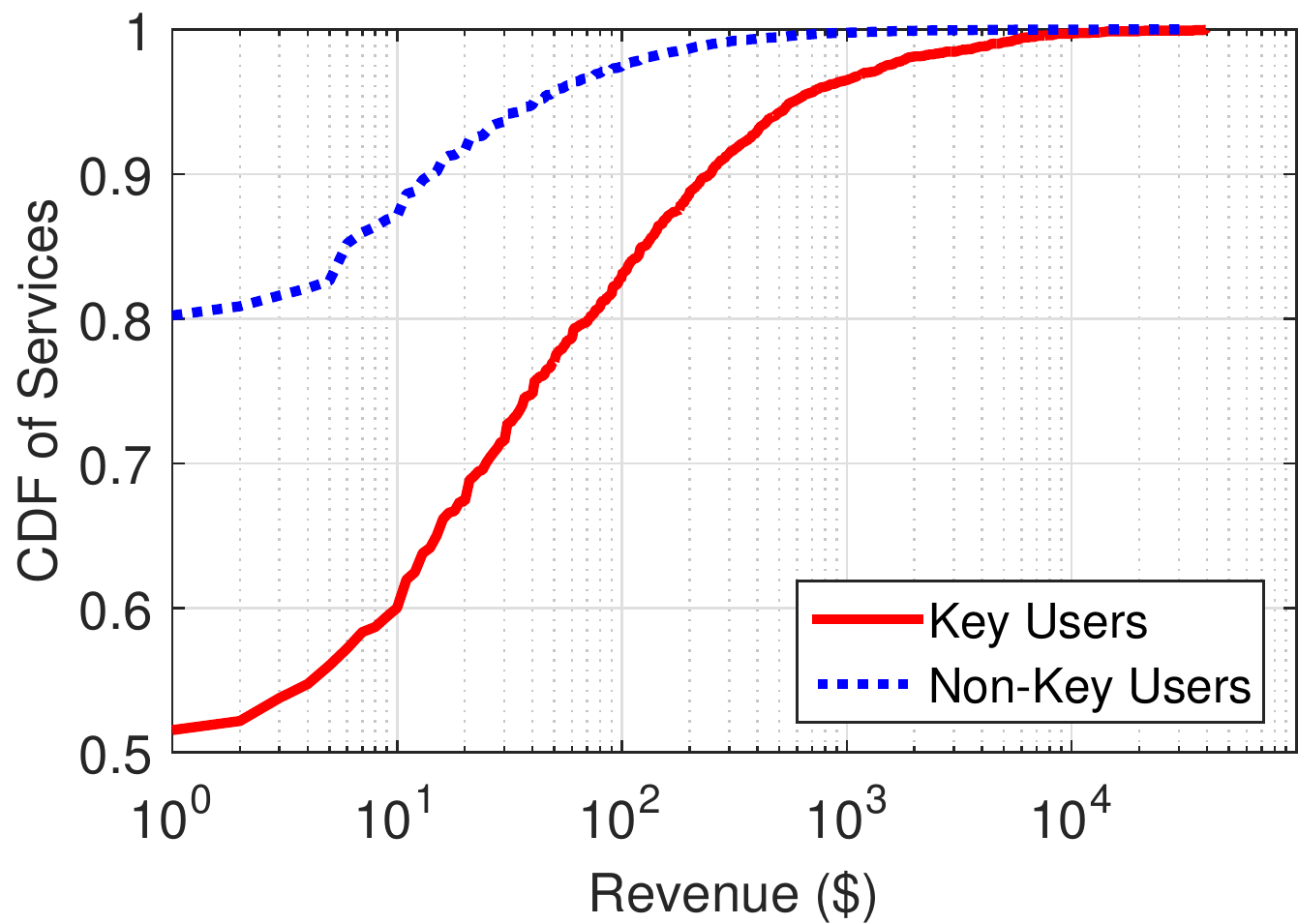}
    \label{fig: service revenue}
    }
    \subfigure[Views]{
    \includegraphics[width=.95\columnwidth]{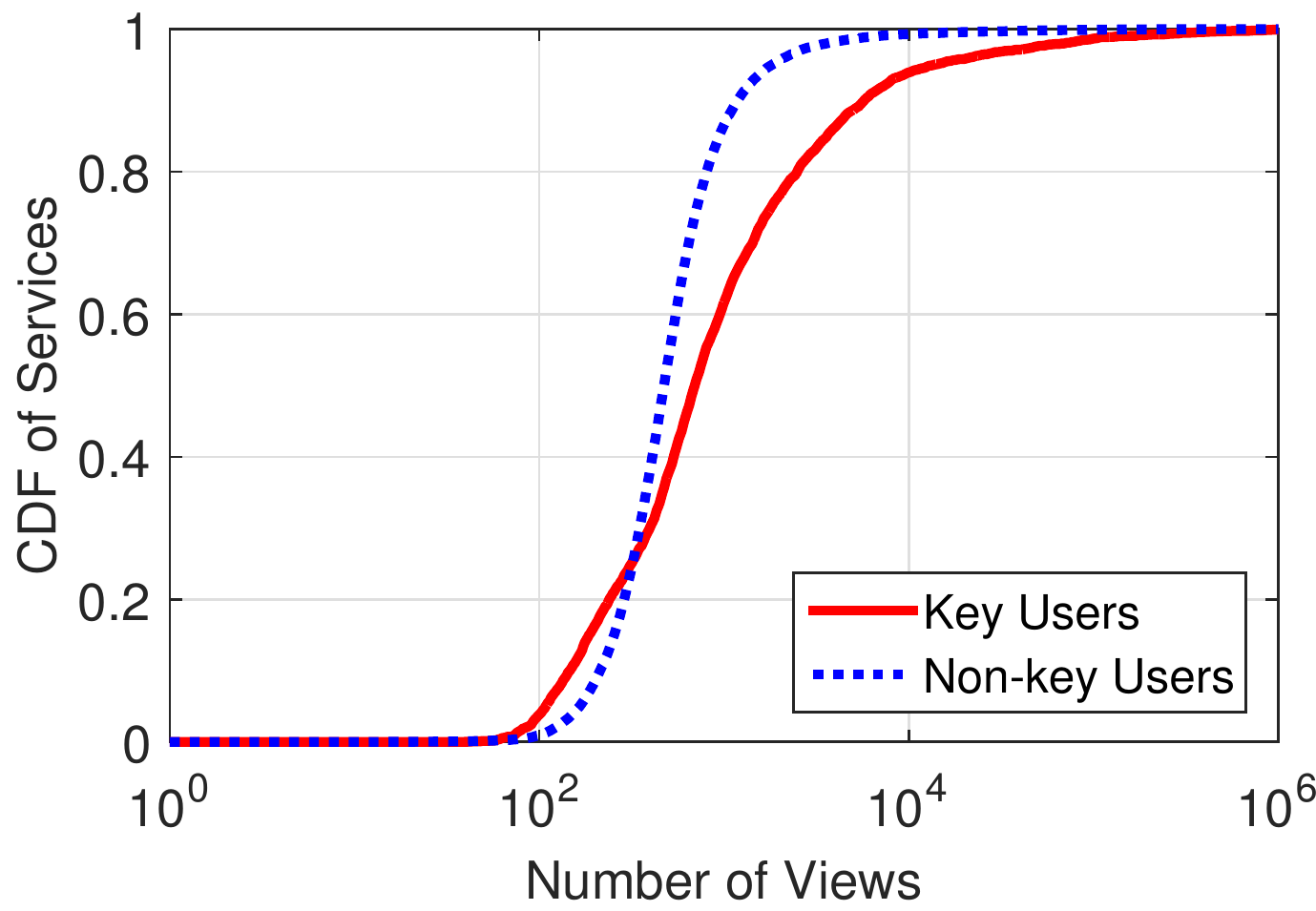}
    \label{fig: service views}
    }
\precaption
\caption{Distribution of service price, volume, revenue, and views on SEOClerks.}
\label{fig: service volume revenue}
\postcaption
\end{figure*}

\descr{Pricing.}
The services on SEOClerks are priced anywhere in the range of \$1-\$999.
Figure \ref{fig: service price} plots the distributions of service prices for key users and non-key users.
We observe that a vast majority of services are priced in the lower range.
For instance, 3,197 (88\%) services offered by key users and 31,719 (80\%) services offered by non-key users are priced up to \$20.
Note that \$999 is the maximum service price allowed by SEOClerks, while \$5 is the minimum allowed service price for the newly registered sellers.
The mode of service price distribution for key users is \$1 and that for non-key users is \$5, which accounts for 416 (11\%) services for key users and 11,227 (31\%) services for non-key users.
As we discuss later, only experienced sellers on SEOClerks are allowed to post services that are priced below the \$5 limit.
Since key users are much more experienced than non-key users, more than a quarter of the services offered by key users are under \$5, while only 11\% of the services offered by non-key users are under \$5.

\descr{Sales.}
We recorded a total of 233,638 sales resulting in the estimated revenue of \$1,349,316 on SEOClerks.
Key users account for more than half of the total sales and revenue.
More specifically, key users made 121,923 sales accounting for an estimated revenue of \$758,959 (56\%), while non-key users made 111,715 sales accounting for an estimated revenue of \$590,357 (44\%).
Figures \ref{fig: service volume} and \ref{fig: service revenue} show the distributions of service volume and revenue for key and non-key users.
It is noteworthy that a vast majority of services by key users (1,874 = 51\%) and non-key users (26,547 = 74\%) have no sales and thus zero revenue.

We observe a skewed distribution of sales volume and revenue.
For key users, 9\% of the services had just one sale, 5\% of the services had two sales, and 3\% of services had three sales.
For non-key users, 8\% of the services had just one sale, 3\% of the services had two sales, and 2\% of services had three sales.
On the other hand, a few popular services account for a large fraction of sales.
For key users, the most popular service in terms of sales volume is {\em ``add 2000 to 2500 Youtube views or 600+ INSTAGRAM Followers or 1000 Likes''} (priced at \$2) and has 3,853 sales resulting in \$7,706 revenue.
For non-key users, the most popular service in terms of sales volume is {\em ``400 Facebook Fanpage likes OR 1300 Twitter Marketing OR 1500 INSTAGRAM Marketing''} (priced at \$3) and has 2,968 sales resulting in \$8,904 revenue.
While we note that low priced services tend to have high sale volume, higher priced services still tend to generate more revenue.
For key users, the top service in terms of revenue is {\em ``Backlinks to improve Google search ranking''} (priced at \$29) attracting 1,364 sales yielding \$39,556 in revenue.
For non-key users, the top service in terms of revenue is {\em ``Google X Factor Link Circle For Higher Ranking And Quality Links''} (priced at \$57) attracting 550 sales yielding \$31,350 in revenue.

\begin{figure*}[!t]
\centering
    \subfigure[Key users]{
    \includegraphics[width=.95\columnwidth]{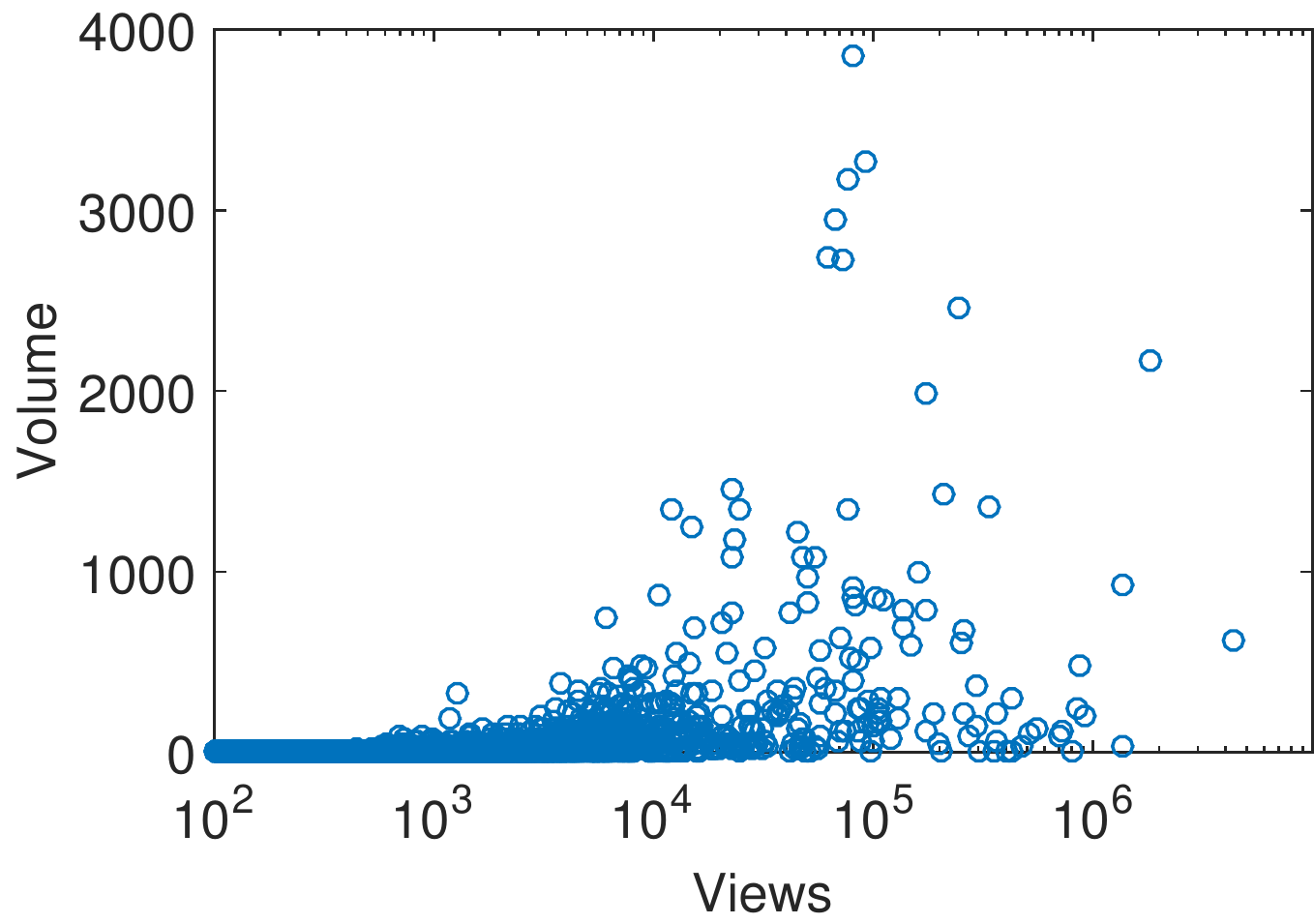}
    \label{fig: scatter service volume view insider}
    }
    \subfigure[Non-key users]{
    \includegraphics[width=.95\columnwidth]{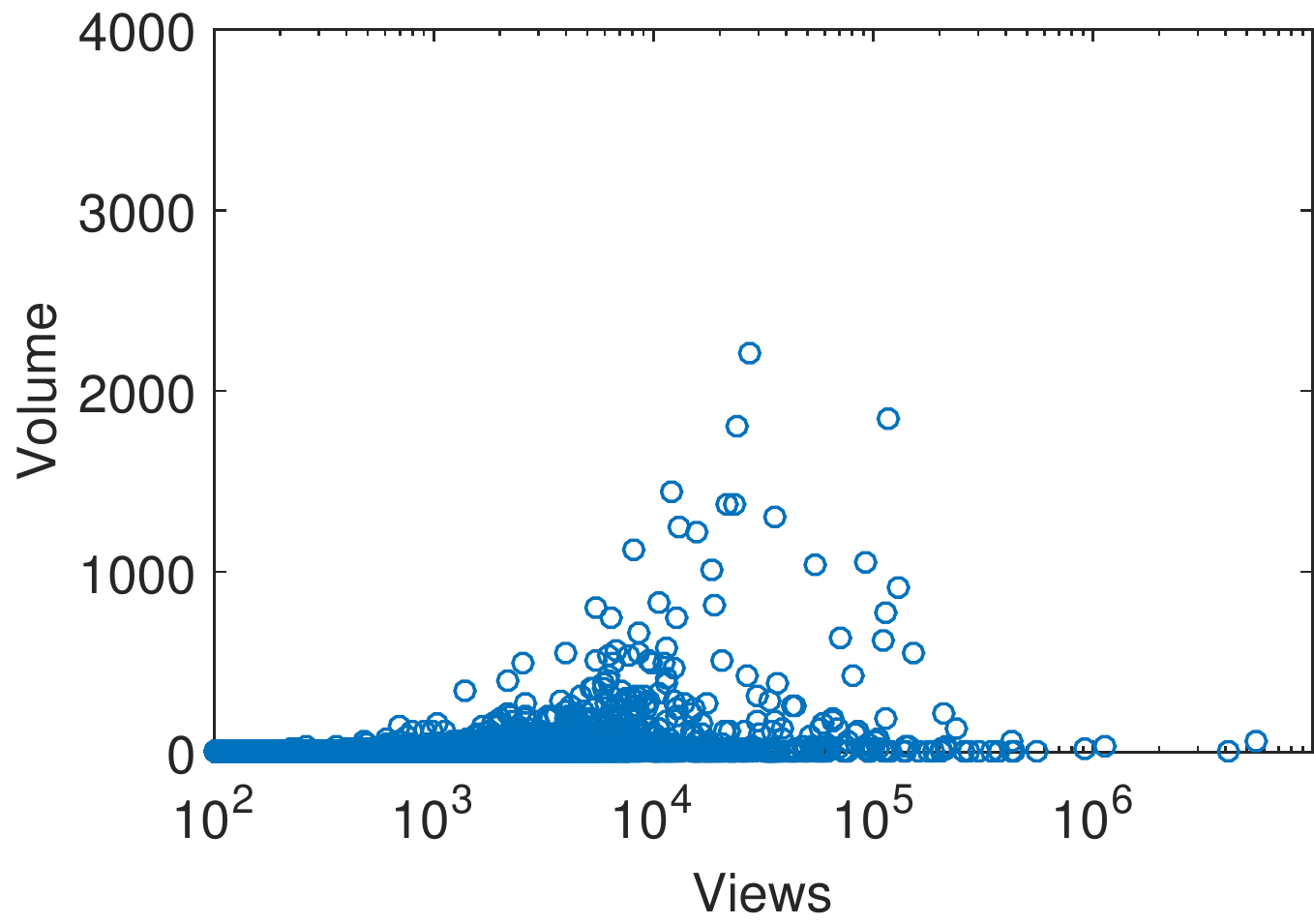}
    \label{fig: scatter service volume view nonInsider}
    }
\precaption
\caption{Scatter plot of service view count and sales volume.}
\label{fig: scatter service volume view}
\postcaption
\end{figure*}

\descr{View Count.}
To further examine why key users account for more sales and revenue, we analyze the correlation between service view count and sales volume.
Figure \ref{fig: service views} plots the distribution of view count for services offered by key users and non-key users.
We note that the services offered by key users are generally viewed more than those by non-key users.
For example, the average number of views for key users is 9,218 while the average number of views for non-key users is 1,962.
Moreover, 1.3\% of the services offered by key users are viewed more than 100 thousand times while only 0.1\% of the services offered by non-key users are viewed over 100 thousand times.
Our eyeball analysis revealed that most services featured on the homepage are posted by key users.
We surmise that the services offered by key users tend to have higher view counts because they are more frequently featured on the marketplace.
To test whether higher view counts translate into more sales, we analyze the correlation between service view count and sale volume.
Figures \ref{fig: scatter service volume view insider} and \ref{fig: scatter service volume view nonInsider} visualize the correlation between service view count and sale volume for key users and non-key users, respectively.
We note that services with more views tend to have higher sales volume for both key and non-key users.
Thus, due to their higher view count, it is expected that the services offered by key users tend to garner more sales than those by non-key users.

\begin{table*}[!t]
\small
\centering
\begin{tabular}{|l|r|r|l|r|r|}
\hline \multicolumn{1}{|c|}{\multirow{2}{*}{\bf{Category}}} & \multicolumn{1}{c|}{\multirow{2}{*}{\bf{\% of Services}}}
& \multicolumn{1}{c|}{\multirow{2}{*}{\bf{Revenue}}} &
\multicolumn{3}{c|}{\bf{Top Service}}\\
\cline{4-6}  &&&  \multicolumn{1}{c|}{\bf{Description}} &  \multicolumn{1}{c|}{\bf{Revenue}} & \multicolumn{1}{c|}{\bf{Price}}  \\
\hline Black-hat SEO & 40.6\% & \$417,865 (55\%) & Backlinks to improve search ranking& \$39,556  & \$29 \\
\hline Instagram & 13.9\% & \$78,274 (10\%) &  1,000 Instagram followers &\$7,706  & \$10 \\
\hline YouTube & 8.0\% & \$73,409 (10\%) &  100,000 safe YouTube views &\$3,5160  & \$120 \\
\hline Twitter & 16.1\% & \$52,583 (7\%) & 50,0000 followers or 2,000 re-tweets &\$4,320   & \$20 \\
\hline Website traffic & 9.5\% & \$49,599 (6\%) & Promote on a large Facebook group &\$8,640   & \$10 \\
\hline
\end{tabular}
\precaption
\caption{Service categories and the most popular service in each category for key users. }
\label{tbl: service category key users}
\end{table*}

\begin{table*}[!t]
\small
\centering
\begin{tabular}{|l|r|r|l|r|r|}
\hline \multicolumn{1}{|c|}{\multirow{2}{*}{\bf{Category}}} & \multicolumn{1}{c|}{\multirow{2}{*}{\bf{\% of Services}}}
& \multicolumn{1}{c|}{\multirow{2}{*}{\bf{Revenue}}} &
\multicolumn{3}{c|}{\bf{Top Service}}\\
\cline{4-6}  &&&  \multicolumn{1}{c|}{\bf{Description}} &  \multicolumn{1}{c|}{\bf{Revenue}} & \multicolumn{1}{c|}{\bf{Price}}  \\
\hline Black-hat SEO & 23.0\% & \$173,081 (29\%) & Rank your website on first page & \$31,350  & \$57 \\
\hline Twitter & 22.2\% & \$82,147 (14\%) &  1 million Twitter followers & \$11,037  & \$849\\
\hline Instagram & 12.6\% & \$47,591 (8\%) &  1,000 Instagram followers &\$4,418  & \$2\\
\hline YouTube & 10.9\% & \$24,417 (4\%) & 8,000 safe YouTube views  &  \$2,052 & \$12 \\
\hline Website traffic & 7.0\% & \$19,226 (3\%) & Views UP - Web Traffic Bot &\$2,360   & \$40\\
\hline
\end{tabular}
\precaption
\caption{Service categories and the most popular service in each category for non-key users. }
\label{tbl: service category non-key users}
\end{table*}

\descr{Service Categorization.}
To systematically analyze different types of fraudulent services on SEOClerks, we use keyword analysis and manual curation to group top selling services into various categories based on their target, e.g., Twitter followers, Instagram followers, search engine manipulation, etc.
Tables~\ref{tbl: service category key users} and ~\ref{tbl: service category non-key users} list the top categories of services and the top selling service for each category for key users and non-key users, respectively.
We note that a majority of services target black-hat search engine optimization and social network reputation manipulation for both key users and non-key users.
Specifically, more than 40\% of services offered by key users target black-hat SEO; whereas, 23\% of services offered by non-key users target black-hat SEO.
Black-hat SEO services account for more than half of the revenue of services sold by key users and 31\% of the total marketplace revenue.
In contrast, more than 50\% of the services offered by non-key users target popular social media platforms while about 28\% percent of services of key users are targeted towards social media platforms.
The largest service category among social media platforms for non-key users is Twitter.
The most popular service in Twitter category provides ``1 million Twitter followers'' for \$849 and has garnered \$11,037 in total revenue.
In contrast, the largest service category among social media platforms for key users is Instagram.
The most popular service in Instagram category provides ``1,000 Instagram followers'' for \$10 and has garnered \$7,706 in total revenue.

\begin{table*}[!t]
\small
\centering
\begin{tabular}{|c|r|c|r|r|c|r|}
\hline \bf{Reputation} &  \multicolumn{3}{c|}{ \bf{Key Users} }  & \multicolumn{3}{c|}{ \bf{Non-Key Users} } \\
\cline{2-7} {\bf Level} & Number of Users & \% of Users & Revenue & Number of Users & \% of Users & Revenue\\
\hline X5  & 1 & 1\% & \$15,560 & 1 & $\approx$0\% & \$599\\
\hline X4  & 2 & 2\% & \$36,680 & 1 & $\approx$0\% & \$259\\
\hline X3  & 25 & 25\% & \$446,643 & 12 & $\approx$0\% & \$54,733\\
\hline 5  & 0 & 0\% & \$0 & 1 & $\approx$0\% & \$99\\
\hline 4  & 1 & 1\% & \$2,351 & 2 & $\approx$0\% & \$20,694\\
\hline 3  & 61 & 62\% & \$217,255 & 960 & 11\% & \$383,626\\
\hline 2  & 0 & 0\% & \$0 & 334 & 4\% & \$12,817\\
\hline 1  & 9 & 9\% & \$40,470 & 7,451 & 85\% & \$117,530\\
\hline
\end{tabular}
\precaption
\caption{User reputation statistics on SEOClerks. }
\label{tbl: user level}
\postcaption
\end{table*}

\subsection{Users}
\label{subsec:users}

\descr{General Stats.}
We find a list of 262,909 users on SEOClerks.
We label a user as a seller if the user has listed at least one service.
Similarly, we label a user as a buyer if the user has purchased at least one service.
Note that a user may be categorized both as seller and buyer.%
We identified 8,861 sellers and 33,092 buyers on SEOClerks.

\descr{Reputation.}
SEOClerks uses a tiered reputation system to categorize users.
The system assigns users one of the available 8 reputation levels.
New users start from level 1.
A user's level is upgraded automatically based on fulfillment of certain requirements for the first five levels (1,2,3,4,5), while level X users (X3,X4,X5) are considered elite and they are selected manually by staff members of SEOClerks.
The details of requirements and benefits for level promotion are described in~\cite{seoclerkslevels}.
A higher reputation level provides more benefits and less restrictions.
For example, users at higher reputation levels can price services below the \$5 limit and get faster payment clearance.

Table~\ref{tbl: user level} lists user reputation level statistics for key users and non-key users on SEOClerks.
We note that key users are generally more experienced than non-key users.
Most key users are at reputation level 3 (62\%) while most non-key users are at reputation level 1.
We surmise that key users receive preferential treatment from the marketplace staff.
For example, we note that 28 out of 99 key users are at reputation level X.
In contrast, only 14 non-key users are at reputation level X even though they contain more than a hundred sellers in the top 10 percentile.

Recall that users can be sellers and/or buyers: in the following, we analyze them separately.

\subsection{Seller Analysis}
\begin{figure*}[t]
\centering
    \subfigure[Services]{
    \includegraphics[width=.95\columnwidth]{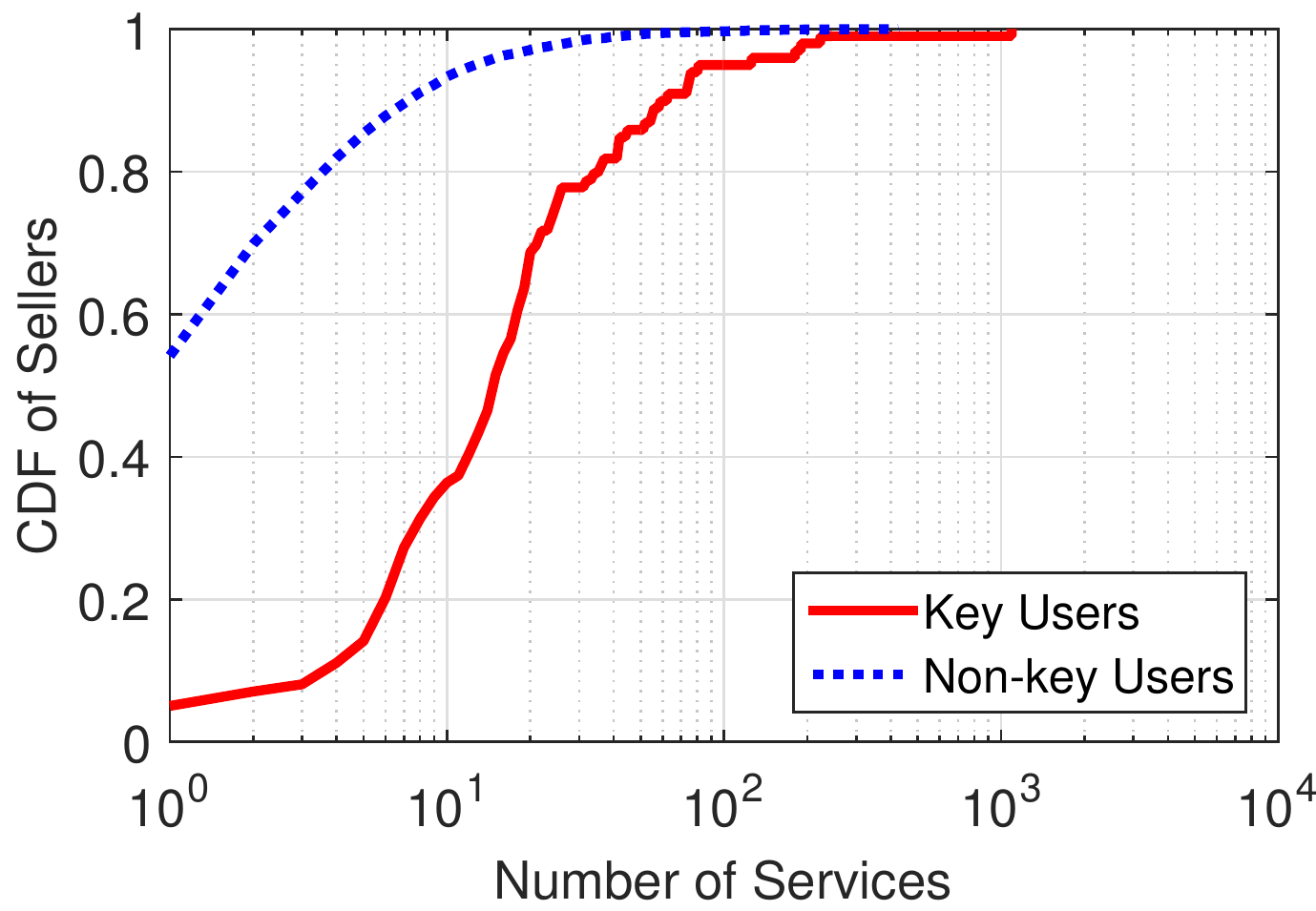}
    \label{fig: number of services}
    }
    \subfigure[Revenue]{
    \includegraphics[width=.95\columnwidth]{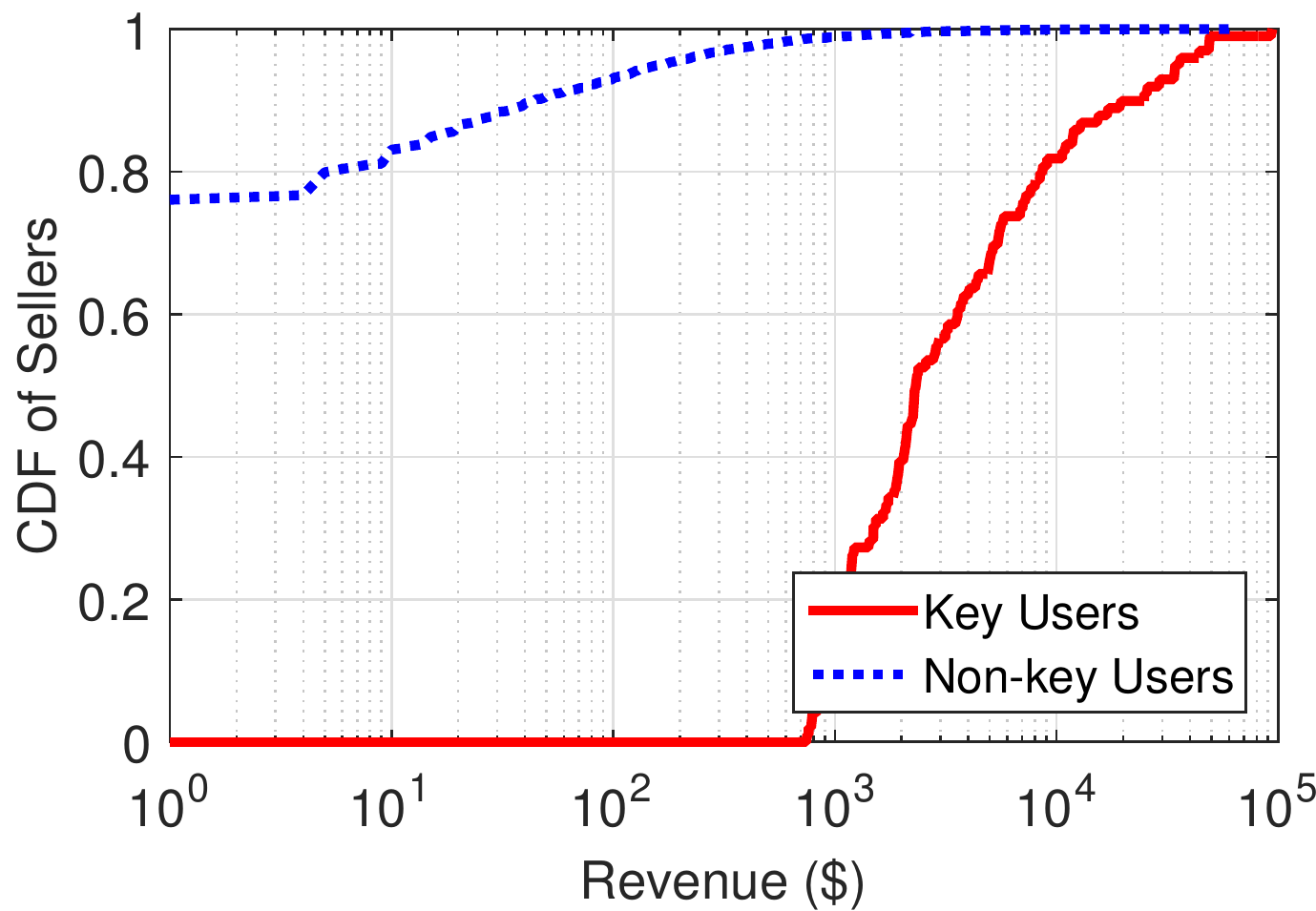}
    \label{fig: seller revenue}
    }
\precaption
\precaption
\caption{Distributions of seller services and revenue on SEOClerks.}
\label{fig: seller stats}
\end{figure*}

We identify 8,861 sellers on SEOClerks, out of which 99 are labeled key users and the remaining 8,762 are labeled as non-key users.
Note that some non-key users have not sold any service yet---these ``zero-sale'' sellers are included in our statistics.

\begin{table}[!t]
\small
\centering
\begin{tabular}{|c|r|r|}

\hline {\bf Country} & {\bf Non-Key Users} & {\bf Key Users} \\
\hline {\bf\em Total}      & {\bf\em 8,762} & {\bf\em 99} \\
\hline India             & 18\% & 29\% \\
\hline USA        & 15\% & 15\% \\
\hline Bangladesh        & 10\% & 18\% \\
\hline Pakistan        & 7\% & 12\% \\
\hline Indonesia        & 5\% & 2\% \\
\hline
\end{tabular}
\precaption
\caption{Geographic location of sellers.}
\label{tbl: geographic distribution}
\postcaption
\end{table}

\descr{Geographic Characteristics.}
SEOClerks provides the geographic location of users based on IP geolocation and/or manual input from users.
Table \ref{tbl: geographic distribution} lists the geographic distribution of sellers across top-five countries.
We note that a substantial fraction of sellers are from a few Asian countries including India, Bangladesh, Pakistan, Indonesia, and Philippines.
This is somewhat expected because of their relatively lower per-capita income \cite{worldbankGDP}.
We also note that key users are concentrated more in Asian countries as compared to non-key users.
Some sellers may be using USA-based VPNs/proxies to manipulate their geolocation for credibility \cite{Muir:2009}.

\descr{Number of Services.}
Figure~\ref{fig: number of services} plots the distribution of the number of services listed by key and non-key users on SEOClerks.
Key users list 3,645 services while the remaining 35,875 services are offered by non-key users.
Note that more than 50\% of non-key users listed only one service and more than 90\% percent posted less than 10 services.
Key users tend to post more services (per seller) as compared to non-key users.
Only 5\% key users posted one service and 64\%  posted more than 10 services.
The seller with most listed services among key users had 1,092 services.
In contrast, the seller with most listed services among non-key users had 458 services.

\descr{Revenue.}
Figure~\ref{fig: seller revenue} plots the distribution of seller revenue for key and non-key users.
Overall, key users account for \$758,959 (56\%) revenue, while non-key users account for \$590,357 (44\%) revenue.
It is noteworthy that more than 75\% of non-key users are zero-sale sellers.
The long-tail distribution indicates that a few sellers account for most revenue for non-key users.
Recall that we labeled top 10\% (228) sellers in terms of revenue as top sellers.
Out of the 228 top sellers, 99 sellers were identified as key users.
The minimum and maximum revenue earned by a key user is \$721 and \$94,190, respectively.
The remaining 129 out of 228 top sellers account for 72\% revenue of non-key users.

\subsection{Buyer Analysis}

\begin{table}[!t]
\small
\centering
\begin{tabular}{|c|r|r|}

\hline {\bf Country} & {\bf Non-Key Users} & {\bf Key Users} \\
\hline {\bf\em Total}      &  {\bf\em 33,013} & {\bf\em 79} \\
\hline USA             & 31\% & 13\% \\
\hline Italy        & 6\% & - \\
\hline UK        &  6\% & 1\% \\
\hline India        &  6\% & 28\% \\
\hline Indonesia        &  4\% & 2\% \\
\hline
\end{tabular}
\precaption
\caption{Geographic location of buyers.}
\label{tbl: geographic distribution buyers}
\postcaption
\end{table}

\begin{figure*}[t]
\centering
    \subfigure[Volume]{
    \includegraphics[width=.95\columnwidth]{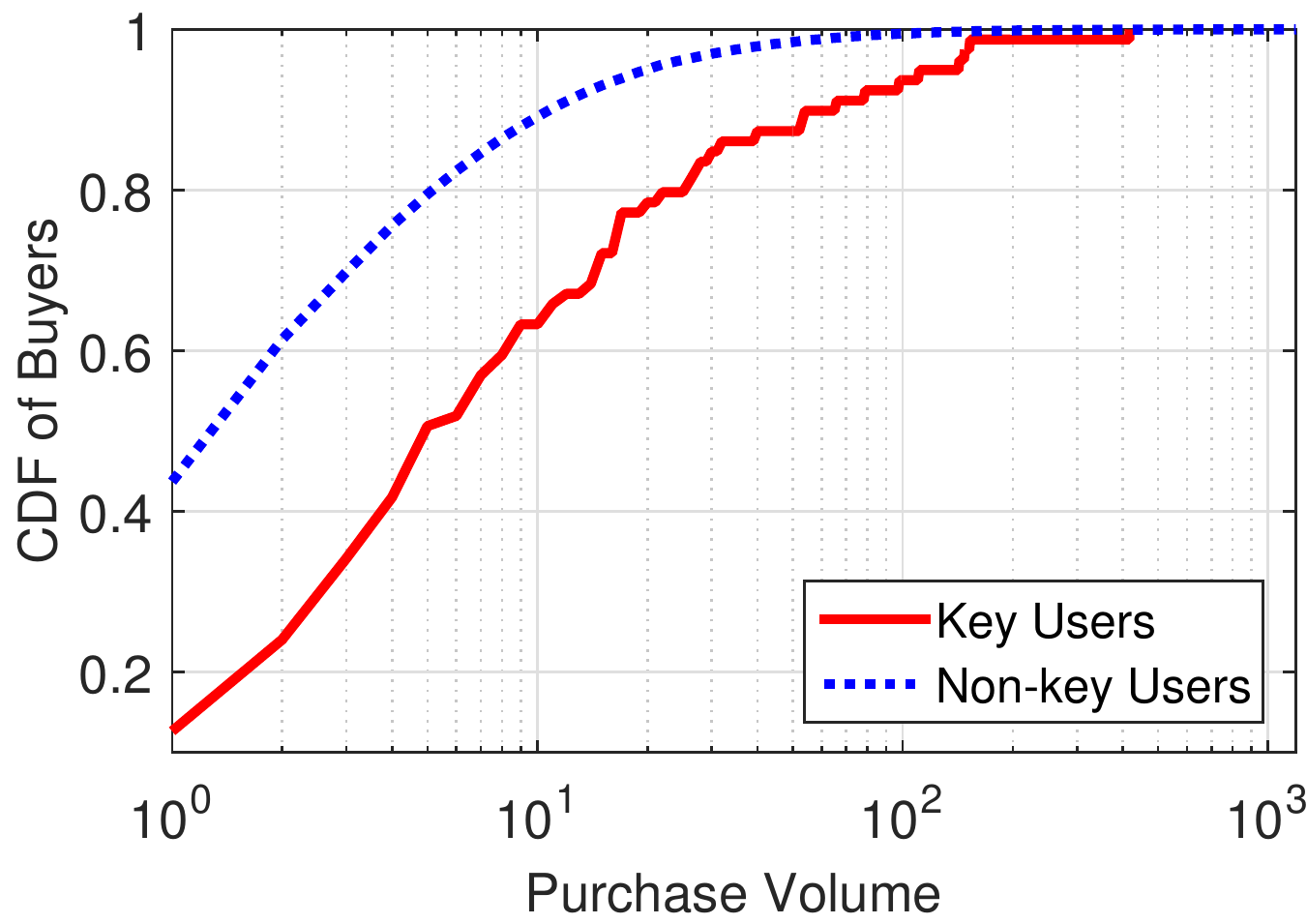}
    \label{fig: buyer volume}
    }
    \subfigure[Expense]{
    \includegraphics[width=.95\columnwidth]{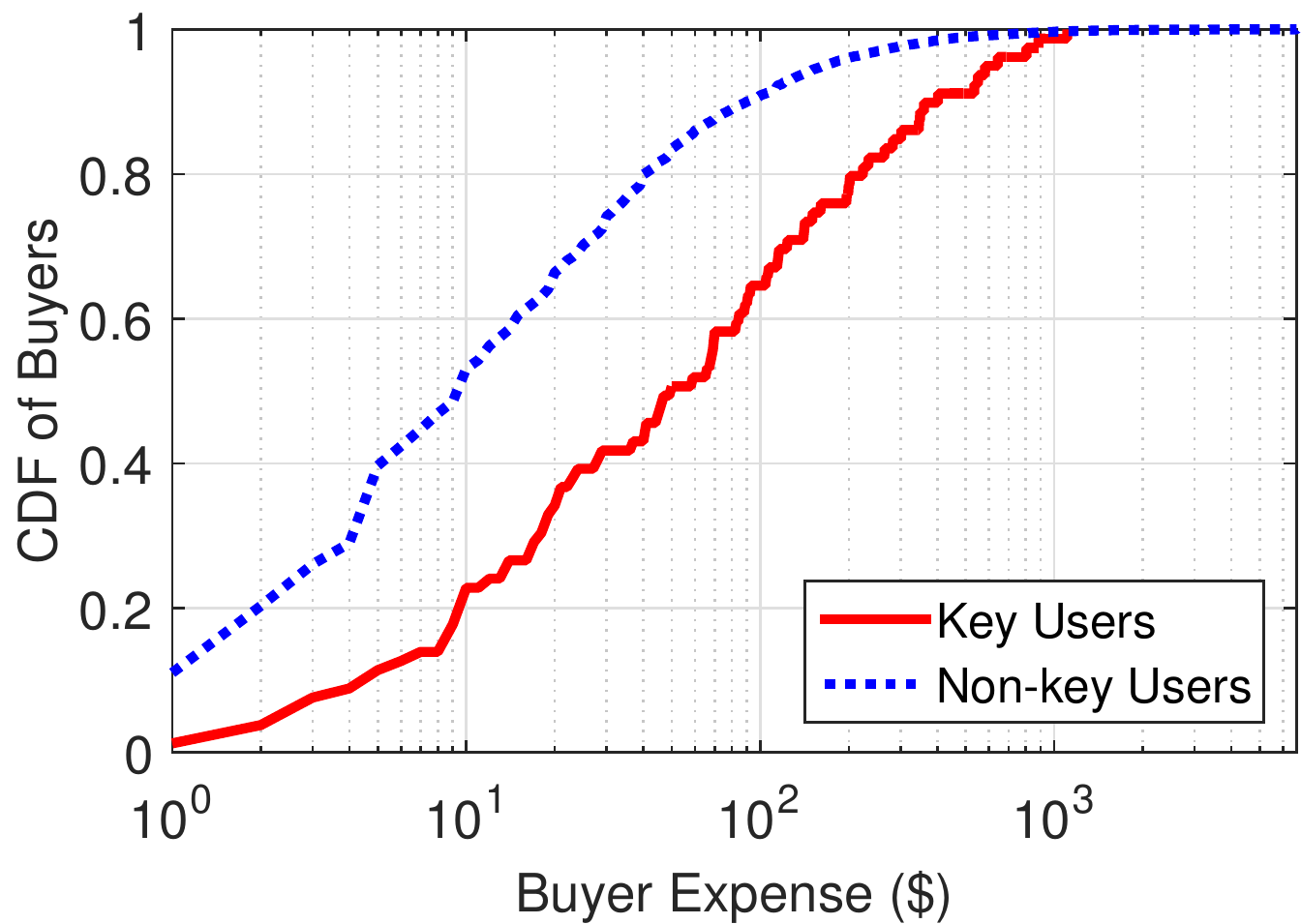}
    \label{fig: buyer expense}
    }\precaption
\precaption
\caption{Distributions of buyer purchase volume and expense on SEOClerks.}
\label{fig: service volume revenue}
\postcaption
\postcaption
\end{figure*}

\begin{figure*}[t]
\centering
    \subfigure[Key Users]{
    \includegraphics[width=.95\columnwidth]{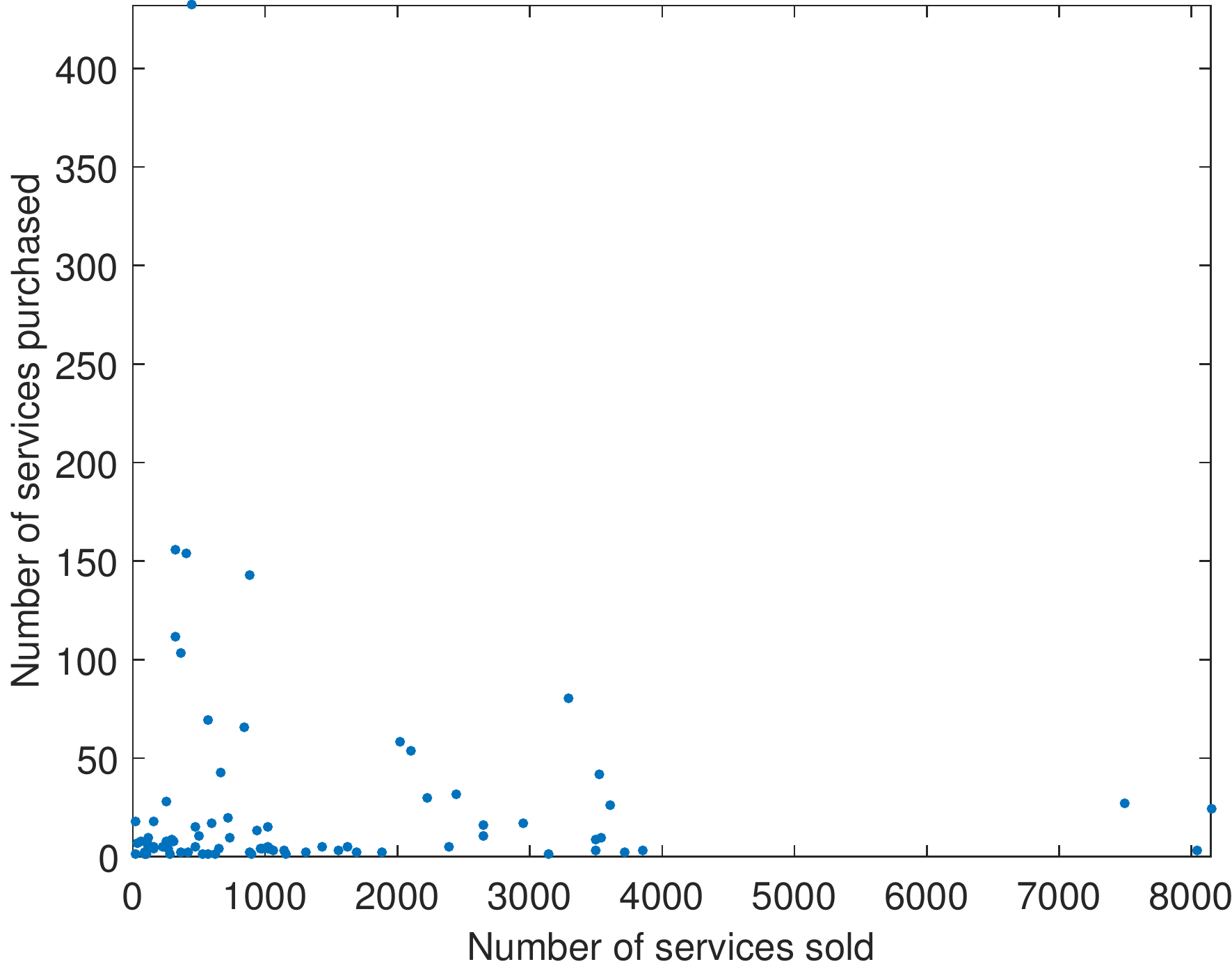}
    \label{fig: KeyUserBuyVsSell}
    }
    \subfigure[Non-key Users]{
    \includegraphics[width=.95\columnwidth]{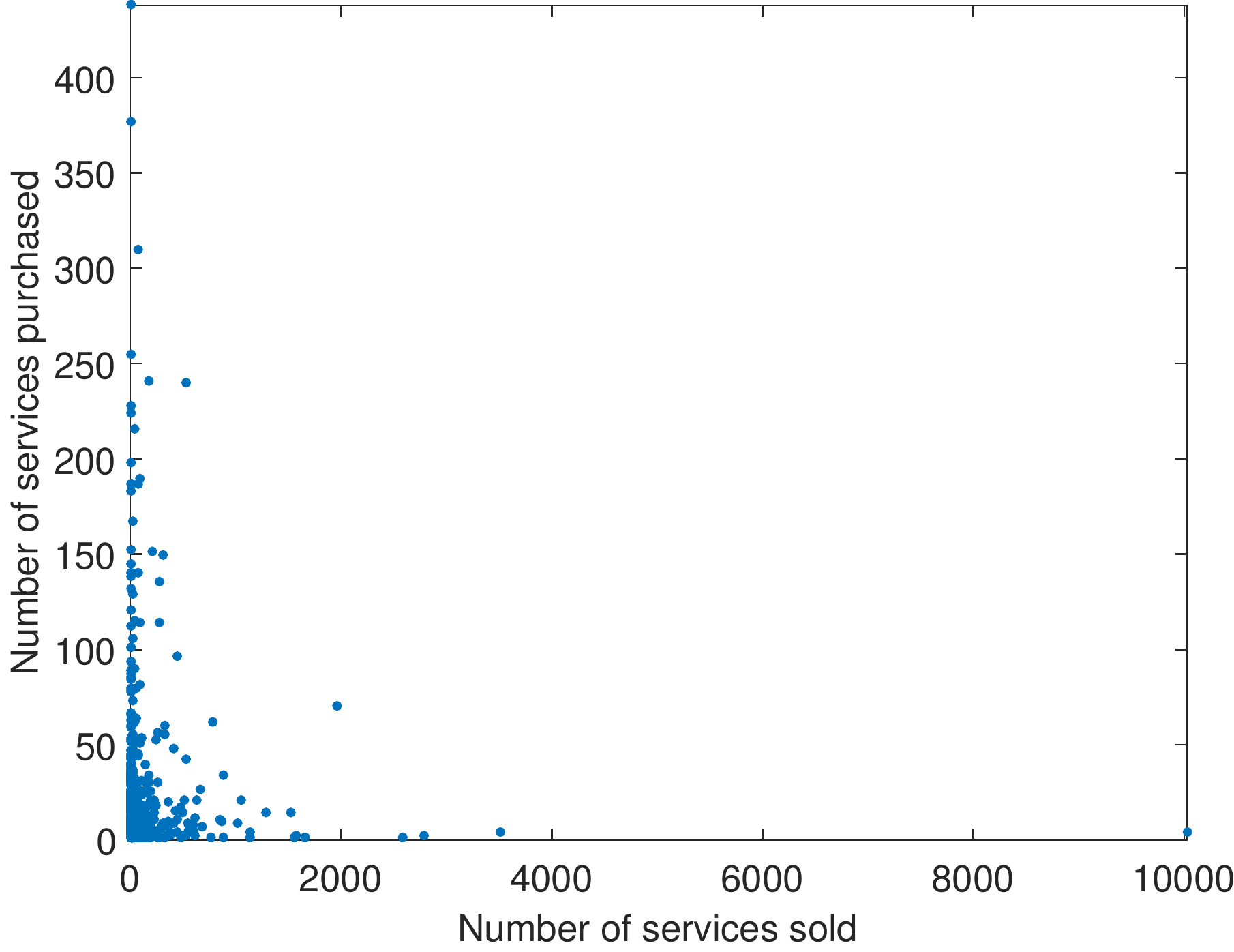}
    \label{fig: NonKeyUserBuyVsSell}
    }\precaption
\precaption
\caption{Each point in the scatterplots represent the number of services sold and purchased by a user on SEOClerks. There are many sellers who are also frequently buying a large number of services}
\label{fig: Buy vs Sell}
\postcaption
\postcaption
\end{figure*}

We identify 33,092 buyers on SEOClerks, out of which 79 buyers are labeled key users and the remaining 33,013 are labeled non-key users.

\descr{Geographic Characteristics.}
Table \ref{tbl: geographic distribution buyers} lists the geographic distribution of buyers across top-five countries.
Overall, buyers are relatively concentrated in the North American and European countries such as USA, Italy, UK, and Canada.
However, we note that a large number of buyers labeled as key users are located in India.
Recall that all buyers who are key users are also top sellers on the marketplace.
These key users also purchase services of other sellers.
Regardless of the role of the marketplace users, our findings somewhat mirror the site's audience statistics as estimated by Alexa.
Alexa estimates that 13.8\% of the site's visitors are from USA, followed by 13.5\% from India, and 4.7\% from Italy.

\descr{Purchase Statistics.}
Figure~\ref{fig: buyer volume} plots the distributions of the purchase volume by key and non-key users.
We note that a majority of key users (88\%) are buyers and they purchased services more than non-key users.
For key users, the median purchase volume is 5 and the average is 24.
For non-key users, the median purchase volume is 2 and the average is 5.
Figure~\ref{fig: buyer expense} plots the distributions of buyer expense (the total amount of money spent by a buyer) by key and non-key users.
We note that key users also spend more money to purchase services as compared to non-key users.
For key users, the median buyer expense is \$50 and average is \$141.
For non-key users, the median buyer expense is \$10 and average is \$41.

\descr{Reselling Behavior.}
We next analyze users with dual roles of a buyer and seller (i.e., they sold at least one service and also purchased at lease one service).
Figure~\ref{fig: Buy vs Sell} visualizes the scatter plot of the services sold and purchased by all dual role key and non-key users on the marketplace.
79 key users and 1,101 non-key users have a dual role of buyers and sellers.
For example, a key user purchased 432 services and also sold 450 services while another non-key user purchased 240 services and sold 530 services.
To understand the behavior of these users, we manually analyze the services purchased and sold by them.
We find that a majority of the dual users are buying and then selling the same kind of services.
This behavior is sometimes due to users purchasing services from other sellers for less price and reselling them at higher prices.
For example, a key user offers a service providing 1,000 Instagram followers for \$4, and the same user has repeatedly purchased similar services from multiple users for \$2.
As another example, a key user offers a service providing 1,000 SoundCloud plays for \$1, and the same user has repeatedly purchased a service from another user providing 15,000 SoundCloud plays for \$1.
We surmise that a user may also sometimes purchase services from other sellers to fulfill existing orders (e.g., due to receiving an unusually large number of orders or temporary infrastructure outages).

\begin{figure*}[t]
\centering
    \subfigure[Services of key users purchased by other key users ]{
    \includegraphics[width=.95\columnwidth]{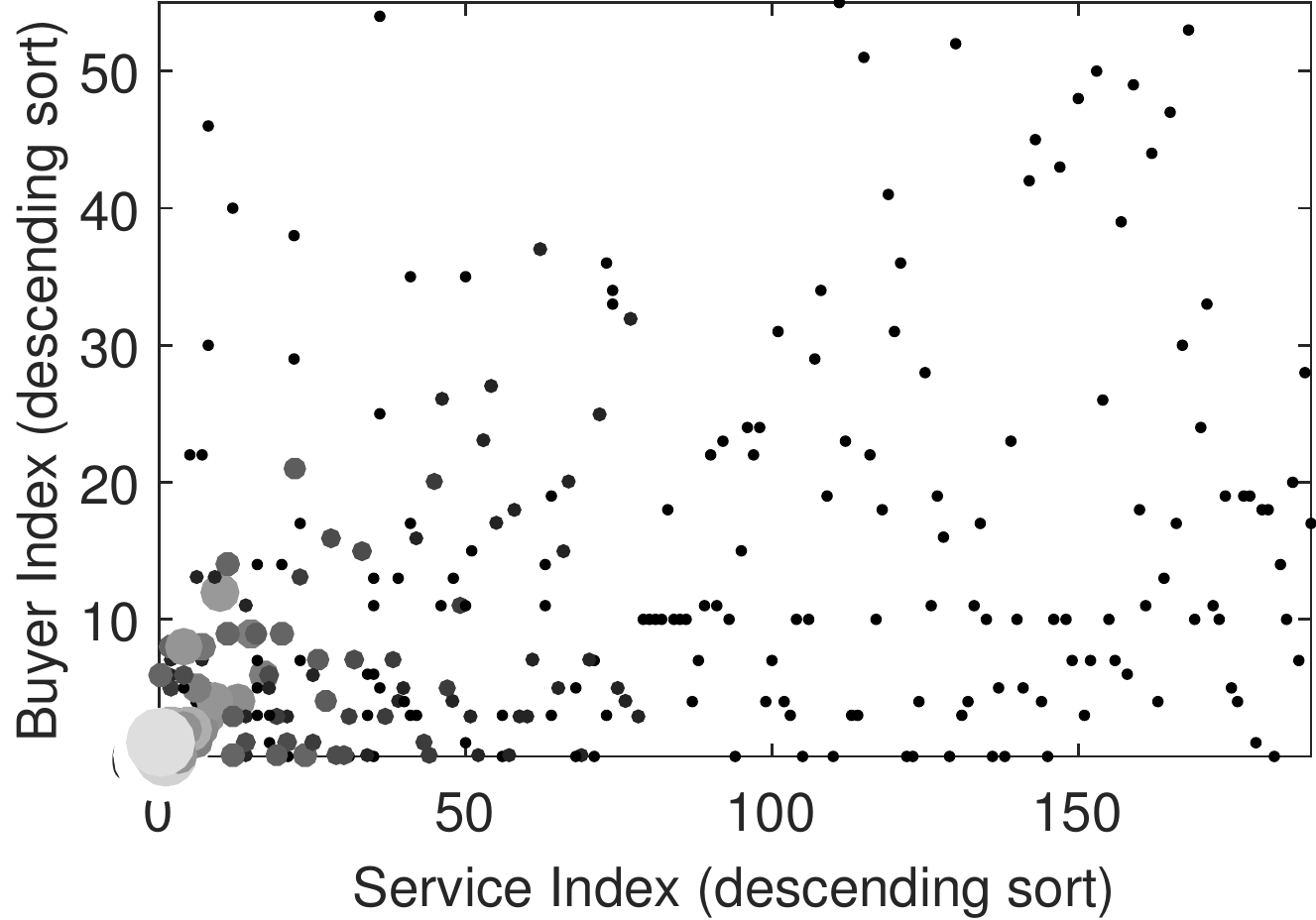}
    \label{fig: buyer service insider insider}
    }
    \subfigure[Services of non-key users purchased by key users]{
    \includegraphics[width=.95\columnwidth]{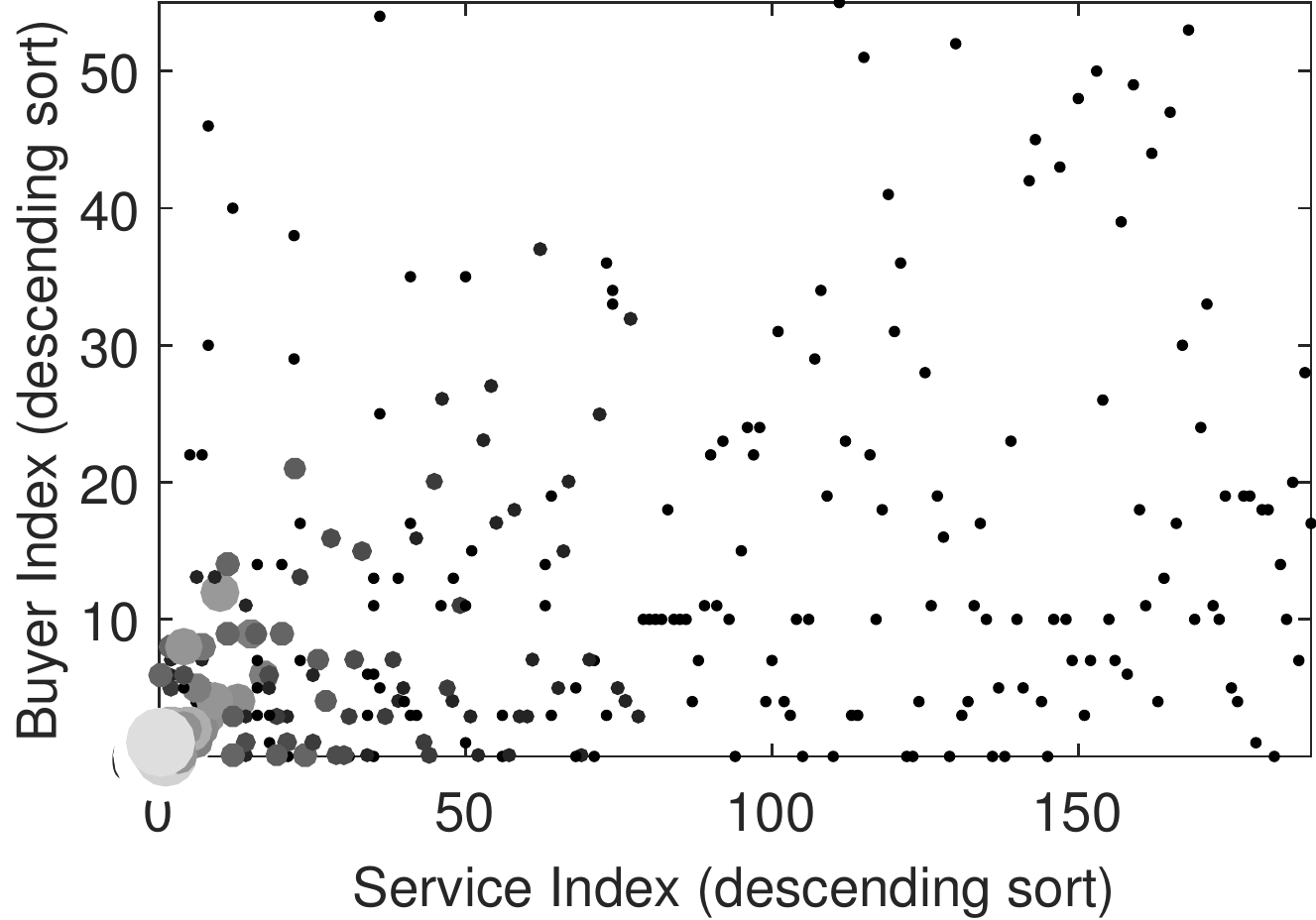}
    \label{fig: buyer service non insider insider}
    }
\subfigure[Services of key users purchased by  non-key users]{
    \includegraphics[width=.95\columnwidth]{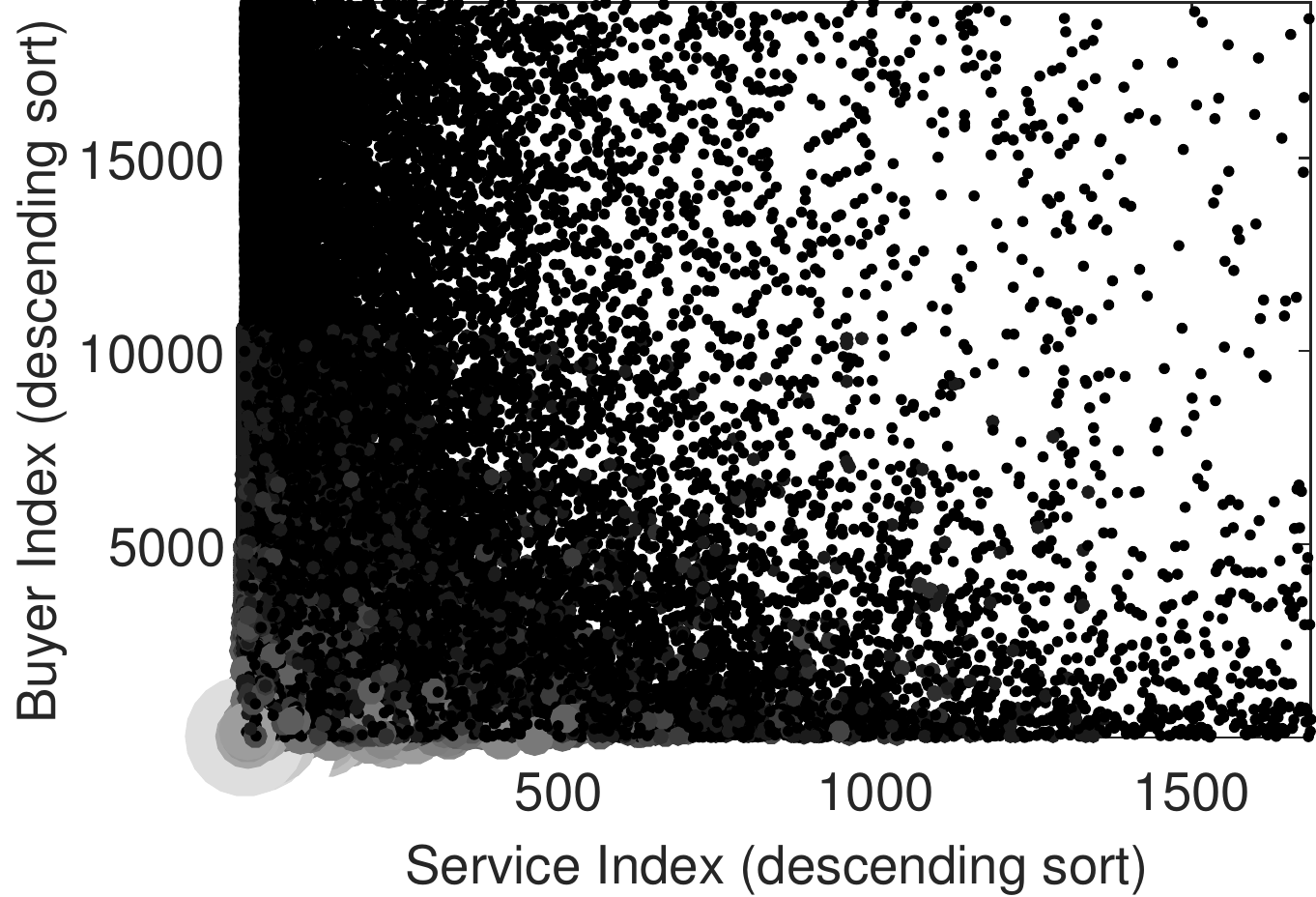}
    \label{fig: buyer service insider nonInsider}
    }
    \subfigure[Services of non-key users purchased by non-key users]{
    \includegraphics[width=.95\columnwidth]{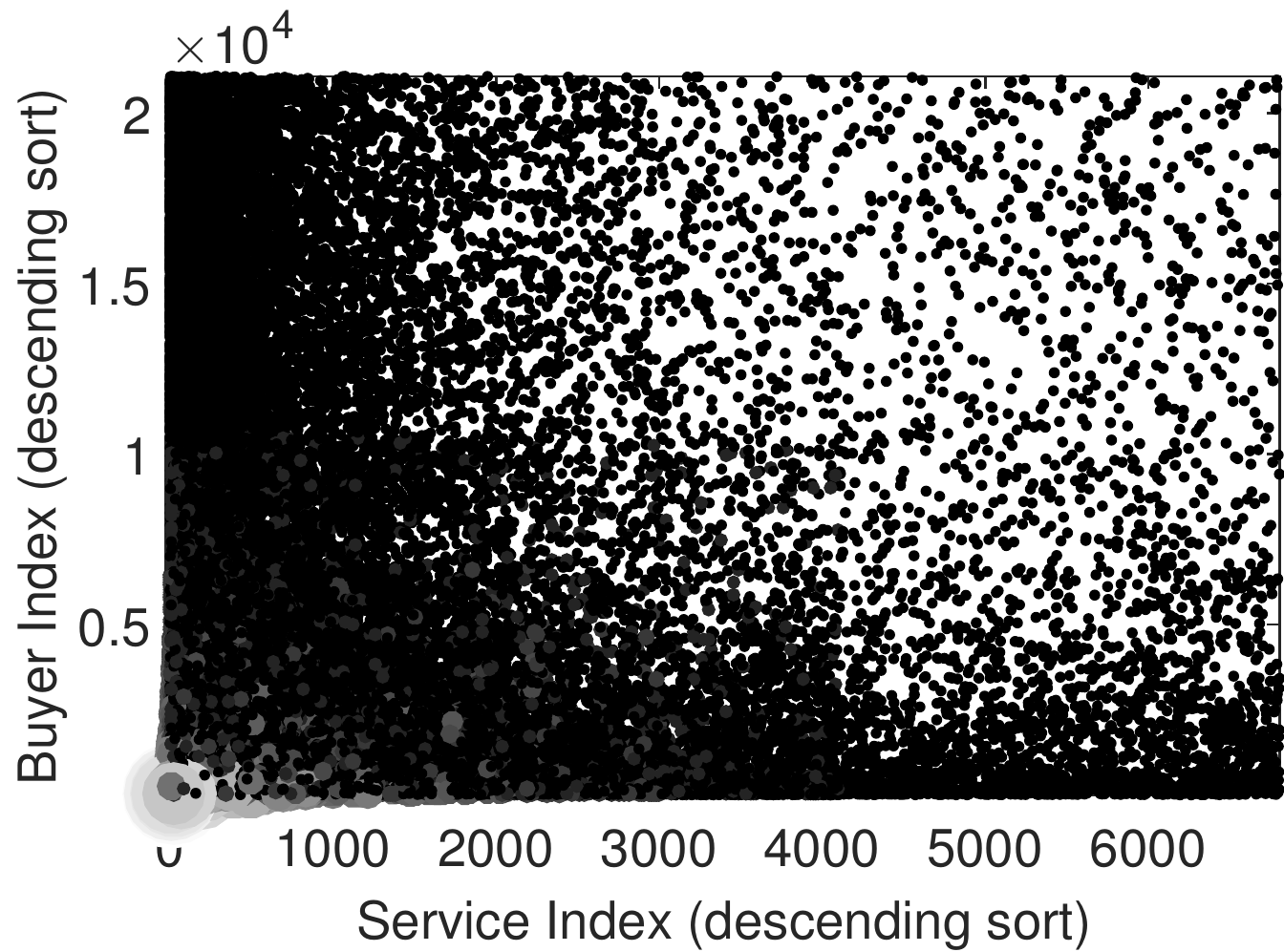}
    \label{fig: buyer service non insider nonInsider}
    }
\precaption
\caption{Scatter plot illustrating the relationship between services and buyers.}
\label{fig: buyer service insider}
\postcaption
\postcaption
\end{figure*}

\descr{Buyer-Service Correlation.}
We next analyze the relationship between buyers and services.
Figure~\ref{fig: buyer service insider} visualizes the scatter plot between buyers and services.
Note that each data point in the scatter plot represents a buyer-service pair, with services and buyers sorted in the descending order with respect to their purchase frequency.
Darker circles represent fewer purchases and lighter circles represent many repeated purchases.
For clarity, we also set the size of circles proportional to purchase frequency.

Figures~\ref{fig: buyer service insider insider} and \ref{fig: buyer service non insider insider} visualize the purchases made by key users from key and non-key users, respectively.
It is interesting to note that many key users have purchased services from other key and non-key users.
Furthermore, a few services tend to have many repeat purchases from several key users (lighter circles are concentrated on the bottom-left of the scatter plot).
To further investigate this finding, we identify the key users who are purchasing a large number of services.
We find that 7 key users are among the top 10\% buyers on the marketplace.
Our manual inspection of these 7 key users revealed that these key users are purchasing services that are similar to the services offered by them.
For example, a key user offers a service providing Instagram followers, and the same user has repeatedly purchased services offering Instagram followers from other sellers.
These are dual role users on the marketplace.

Figures~\ref{fig: buyer service insider nonInsider} and \ref{fig: buyer service non insider nonInsider} visualize the purchases made by non-key users from key users and non-key users, respectively.
We observe that a vast majority of non-key users buy a service only once.
However, a few popular services by key users tend to have many repeat buyers (lighter circles are concentrated on the bottom-left of the scatter plot).

\section{Discussion}

We start the discussion by presenting further interesting observations.

\subsection{Additional Findings}

\descr{Marketplace Commission.} SEOClerks charges 20\% commission for each order. It also charges a nominal transaction processing fee (varying depending on the mode of payment). The 20\% commission is charged from sellers and the transaction processing fee is charged from buyers, thus, according to our estimates, the operators of SEOClerks have earned at least \$269,863 in commissions. Note that SEOClerks operators also offer a variety of services to temporarily ``feature'' services on the marketplace homepage. Based on the number of transactions of these services, we estimate that SEOClerks operators have earned thousands of dollars.

\descr{Revenue Underestimation.} Recall that buyers are not mandated to provide feedback ratings on SEOClerks. Moreover, SEOClerks allows sellers to list ``service extras'' which cost in addition to the base service price. From our crawled data, we cannot tell whether a buyer bought service extras, therefore our revenue estimate represents the lower bound on the actual marketplace revenue.

\descr{Data Trust.} Given the black-hat nature of SEOClerks, it is possible that some information on the websites (e.g., user levels) may be manipulated by the marketplace operators. While we cannot completely rule this out, we created fresh accounts on both marketplaces and positively verified their information (e.g., geographic location, join date, user level, etc.) to lend some confidence to our collected data.

\subsection{Countermeasures}
We now discuss potential countermeasures to curb the activities of black-hat marketplaces, including those targeting key users on SEOClerks as well as
the operators.

\descr{Targeting Key Users.} While key users constitute less than 0.04\% of SEOClerks users, they account for more than half of the revenue. Specifically targeting these key users can considerably limit fraudulent activities out of black-hat marketplaces. Furthermore, active experiments could be conducted to understand the working of their infrastructure, e.g., creating honeypot accounts to identify the fake accounts used for providing likes/followers \cite{stringhini13twitterfollower,decristofaro14facebooklikefarms}.

\descr{Targeting Marketplace Operators.} Another approach is to go after the monetary systems used by black-hat marketplaces. More specifically, SEOClerks uses an escrow mechanism to get transaction/commission fees and to resolve disputes between sellers and buyers. Buyers on SEOClerks can purchase services using standard credit/debit card, PayPal, Payza, or using cryptocurrencies. For PayPal and Payza, the marketplace account of SEOClerks is registered to Ionicware Inc. For all cryptocurrency transactions, SEOClerks uses an account on BitPay which is also registered to Ionicware Inc. These marketplace accounts on PayPal, Payza, and BitPay can be targeted for economic and legal interventions. Another possible countermeasure would be to seek court injunctions and shutdown these websites~\cite{facebooksuit,amazonsuit} by targeting either the domain registrar or the hosting company. However, this action might not be as effective due to possibly lengthy procedures, possibly allowing websites to change name and/or relocate to other hosting providers.

\section{Related Work} \label{sec: related work}
Prior work has looked at black-hat marketplaces to analyze them in terms of demographics, nature and quality of offered services, revenue models, and financial intervention.
While our analysis is also based on measurements (e.g., via periodic crawls) as in some of the related work, there are two key differences between this paper's methodology and prior work: (i) the object of our measurement campaign, and (ii) our investigation aimed to identify key stakeholders who dominate the black-hat marketplace.
To the best of our knowledge, we present first-of-its-kind study to identify and understand the role of key stakeholders on black-hat marketplaces.

\descr{Crowdturfing markets.} Wang et al. \cite{wang12crowdturfingWWW} studied ``crowdturfing'' (astroturfing + crowdsourcing) on two large Chinese malicious crowdsourcing markets (Zhubajie and Sandaha), and surveyed several USA-based and Indian malicious crowdsourcing sites such as ShortTask, MinuteWorkers, etc. Unlike our work, they focused on buyer-driven malicious crowdsourcing markets. Overall, in addition to the market size estimation, they were able to measure real-world ramifications of these services by becoming active customers in one of these markets. Xu et al. \cite{Xu:2015:ERM:} analyzed several black-hat marketplaces. They found that, compared to normal sellers, fraudulent sellers escalate their reputations at least 10 times. Thus, fraudulent sellers profit by harnessing crowd-sourced human laborers to conduct fake transactions for their offered services. Note that SEOClerks is ranked higher than these marketplaces and---unlike most other black-hat marketplaces---provides non-anonymized transaction information which allows us to analyze selling and buying behavior of users in detail. In \cite{motoyama11dirtyjobs} and \cite{lee14darksidemicrotask,lee15turfingfivertwitter}, the authors studied services and crowdturfing, respectively, on Freelancer and Fiverr. They developed machine learning models to detect crowdturfing within mostly legitimate content. Our work confirms many findings from \cite{motoyama11dirtyjobs,lee14darksidemicrotask,ge15gigssupersellerscrowdsourcing,lee15turfingfivertwitter} in terms of services popularity and target. However, our analysis differs in that Fiverr and Freelancer offer mostly legitimate services (more than 80\%, according to the authors), whereas SEOClerks is a dedicated black-hat marketplace.

\descr{Standalone merchants.} Thomas et al. \cite{thomas13traffickingfraudtwitteraccounts} analyzed trafficking of fake accounts in Twitter. They bought accounts from 27 merchants and developed a classifier to detect them. Based on this classifier, they successfully identified several million fraudulent accounts, of which 95\% were disabled with the help of Twitter. In a similar study, Stringhini et al. \cite{stringhini12twitterfollowermarketWOSN,stringhini13twitterfollower} measured the market of Twitter followers, providing Twitter followers for sale. Based on this measurement campaign, the authors evaluated several machine learning techniques to detect sybil accounts. In our prior work \cite{decristofaro14facebooklikefarms}, we presented a measurement study of Facebook like farms, which provide paid services to boost the number of page likes. We note that this line of research focuses on individual merchants and their operational aspects, whereas our work studies operation of black-hat marketplaces involving thousands of merchants.

\descr{Underground forums and markets.} Motoyama et al. \cite{motoyama11undergroundforums} analyzed social dynamics in six underground forums and categorized illegal merchandize traded on these forums. Christin \cite{christin13silkroad} studied Silk Road, an anonymous underground marketplace for contrabands, drugs, and pornography, providing a detailed analysis of the items being sold and the seller population. Buyer feedback was used to estimate total revenue and volume of the transactions. Silk Road data suggests a core clique of top sellers, and our analysis shows a similar trend, where a small group of sellers joined the marketplace early and also happen to be the most successful sellers. Soska et al. \cite{Soska2015MeasuringAnonymousUSENIX} conducted a longitudinal analysis of 16 underground online marketplaces over a time period of two and a half years to understand the evolution of online anonymous marketplaces. These anonymous marketplaces do not expose individual buyer information, thus the authors were unable to perform analysis of buyers.

\section{Conclusion} \label{sec: conclusions}
This paper presented a comprehensive analysis of key stakeholders in a popular online black-hat marketplace, \url{SEOClerks.com}.
These \emph{key users} are among the early joiners, are most active, and make the most money on the marketplace.
Specifically, 99 key users (out of a total of 262K users) account for more than 56\% of the total revenue.
We compare and contrast key and non-key users by analyzing the services they offer, and their selling and buying behavior.
We find that a majority of key users on SEOClerks are located in Asian countries, and that some  of them purchase services from other sellers and then sell them at higher prices.

Black-hat marketplaces constitute a key link in the Internet fraud chain \cite{levchenko11clicktrajectories}.
Overall, our findings highlight opportunities for economic and legal intervention to counter black-hat marketplaces, as we demonstrate that a significant part of the activity is concentrated in the hands of relatively few actors.
More specifically, since key users constitute a majority of the marketplace revenue, targeting them specifically can considerably limit fraudulent activities out of black-hat marketplaces.
In future, we are interested in studying the role of key users on multiple black-hat marketplaces over time. In the long term, our goal is to develop statistical models for early detection of key users to minimize activities out of black-hat marketplaces.

\bibliographystyle{abbrv}

\end{document}